\documentclass[useAMS,usegraphicx,usenatbib]{mn2e} 
\usepackage{graphicx}
\usepackage{graphicx,float}
\usepackage{latexsym,amsmath,amssymb}
\usepackage{natbib}
\usepackage{times}
\usepackage{changebar}
\usepackage{subfigure}
\usepackage{ltxtable}
\usepackage{txfonts}
\usepackage{longtable, lscape}

\usepackage{natbib}
\bibpunct{(}{)}{;}{a}{}{,}

\title[Iron K$\alpha$ emission in Sy1s and Sy2s]{Iron K$\alpha$ emission in type-I and type-II Active Galactic Nuclei}
\author[C. Ricci et al.]{C. Ricci$^{1}$\thanks{E-mail:
ricci@kusastro.kyoto-u.ac.jp},  Y. Ueda$^{1}$, S. Paltani$^{2}$, K. Ichikawa$^{1}$, P. Gandhi$^{3}$,  and H. Awaki$^{4}$\\
$^{1}$Department of Astronomy, Kyoto University, Oiwake-cho, Sakyo-ku, Kyoto 606-8502\\
$^{2}$Department of Astronomy, University of Geneva, ch. d'Ecogia 16, 1290 Versoix, Switzerland\\
$^{3}$Department of Physics, University of Durham, South Road, Durham DH1 3LE, UK \\
$^{4}$Department of Physics, Ehime University, Matsuyama, 790-8577, Japan }
\begin{document}
\date{Received; accepted}

\pagerange{\pageref{firstpage}--\pageref{lastpage}} \pubyear{2011}

\maketitle

\label{firstpage}

\begin{abstract}

The narrow Fe\,K$\alpha$ line is one of the main signatures of the reprocessing of X-ray radiation from the material surrounding supermassive black holes, and it has been found to be omnipresent in the X-ray spectra of active galactic nuclei (AGN). In this work we study the characteristics of the narrow Fe\,K$\alpha$ line in different types of AGN. Using the results of a large {\it Suzaku} study we find that Seyfert\,2s have on average lower Fe\,K$\alpha$ luminosities than Seyfert\,1s for the same 10--50\,keV continuum luminosity. Simulating dummy Seyfert\,1s and Seyfert\,2s populations using physical torus models of X-ray reflected emission, we find that this difference can be explained by means of different average inclination angles with respect to the torus, as predicted by the unified model. Alternative explanations include differences in the intensities of Compton humps, in the photon index distributions or in the average iron abundances. We show that the ratio between the flux of the broad and narrow Fe K$\alpha$ line in the 6.35--6.45\,keV range depends on the torus geometry considered, and is on average $<25\%$ and $<15\%$ for type\,I and type\,II AGN, respectively.
We find evidence of absorption of the narrow Fe\,K$\alpha$ line flux in Compton-thick AGN, which suggests that part of the reflecting material is obscured. We estimate that on average in obscured AGN the reflected radiation from neutral material is seen through a column density which is $1/4$ of that absorbing the primary X-ray emission. This should be taken into account in synthesis models of the CXB and when studying the luminosity function of heavily obscured AGN. We detect the first evidence of the X-ray Baldwin effect in Seyfert\,2s, with the same slope as that found for Seyfert\,1s, which suggests that the mechanism responsible for the decrease of the equivalent width with the continuum luminosity is the same in the two classes of objects.

\end{abstract}

  \begin{keywords}
Galaxies: Seyferts -- X-rays: galaxies -- Galaxies: active -- Galaxies: nuclei 
               
\end{keywords}

\begin{figure*}
\centering
\begin{minipage}[!b]{.48\textwidth}
\centering
\includegraphics[width=9cm]{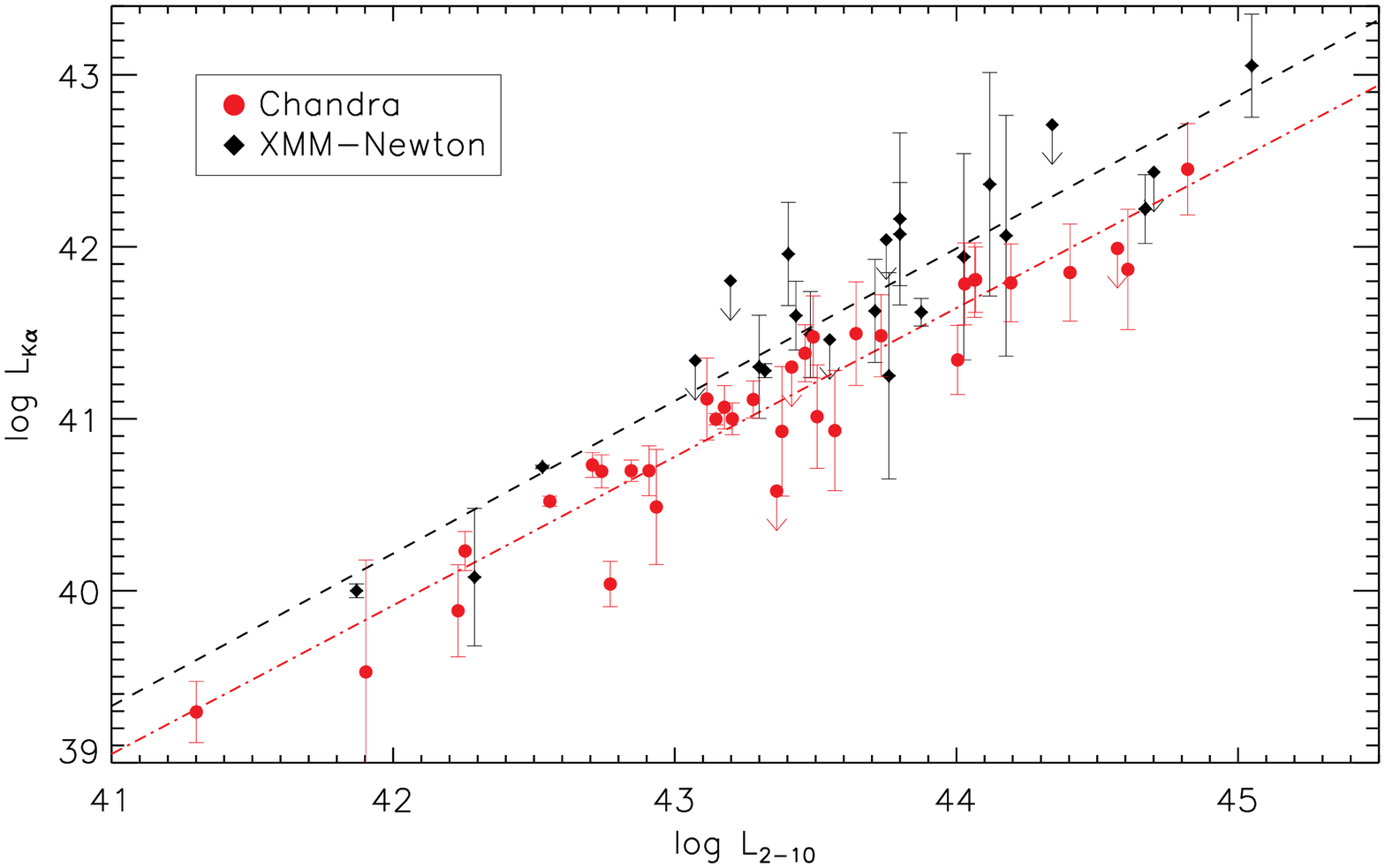}
\end{minipage}
\hspace{0.05cm}
\begin{minipage}[!b]{.48\textwidth}
\centering
\includegraphics[width=9cm]{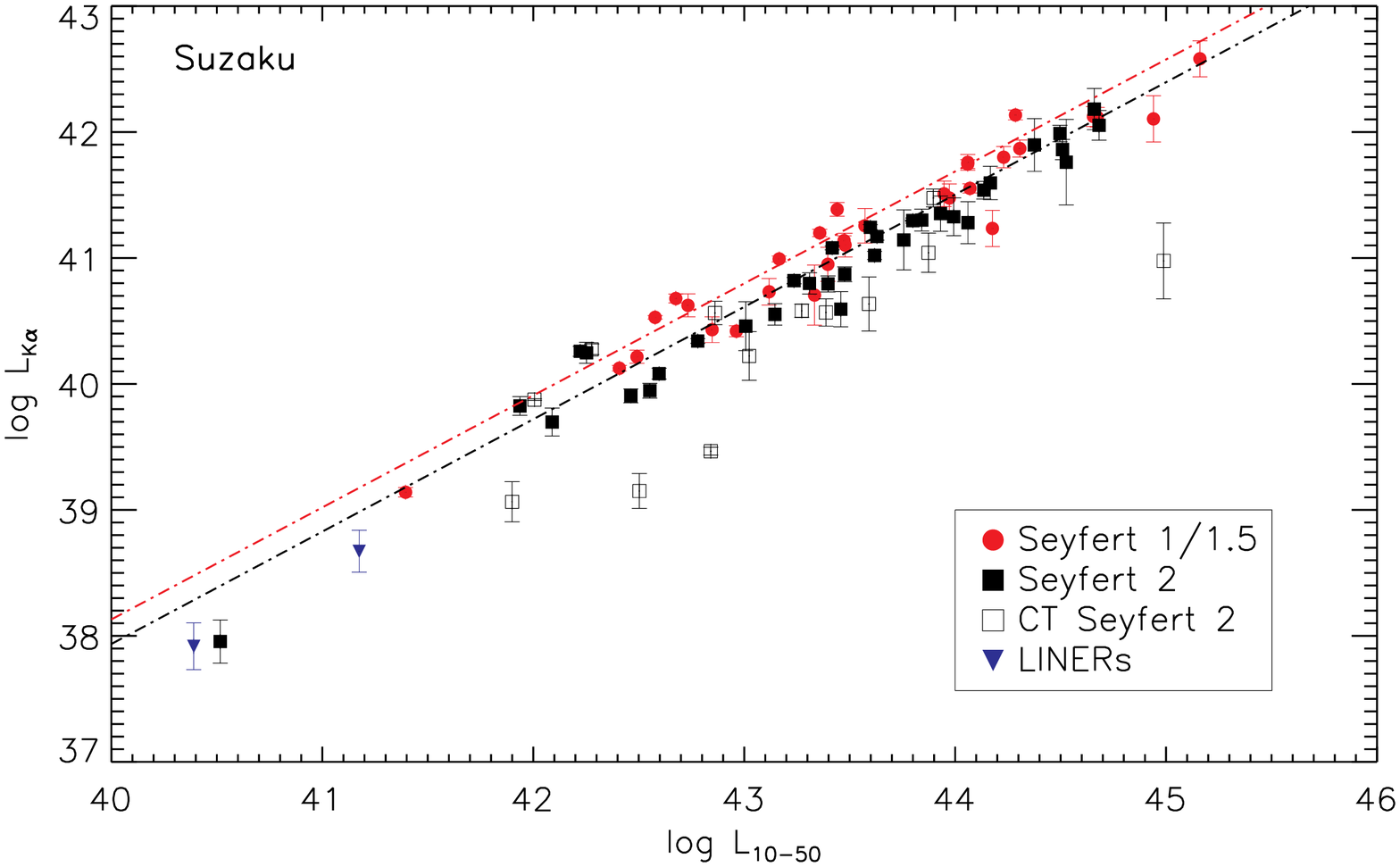}\end{minipage}
 \begin{minipage}[t]{1\textwidth}
  \caption{{\it Left panel}: Scatter plot of the iron K$\alpha$ luminosities versus the 2--10\,keV continuum luminosity of Sy1/1.5s for the {\it XMM-Newton} sample of \citeauthor{Ricci:2014vs} (\citeyear{Ricci:2014vs}, black diamonds) and for the {\it Chandra} sample of \citeauthor{Shu:2010zr} (\citeyear{Shu:2010zr}, red circles). The black dashed and red dot-dashed line represent the best fit obtained applying Eq.\,\ref{eq:Lumirvscont} to the {\it XMM-Newton} and {\it Chandra} sample, respectively. {\it Right panel}: Scatter plot of the iron K$\alpha$ luminosities versus the 10--50\,keV absorption-corrected continuum luminosity for the {\it Suzaku} sample of \citet{Fukazawa:2011fk}. The red and black dot-dashed lines represent the log-linear fit (Eq.\,\ref{eq:Lumirvscont2}) for Seyfert\,1s and Seyfert\,2s, respectively. }
\label{Fig:LcLkalpha}
 \end{minipage}

\end{figure*}

\section{Introduction}

Active Galactic Nuclei (AGN) emit a large fraction of their luminosity in the X-rays. Such an emission is believed to originate from Comptonisation of optical/UV photons produced in the accretion disk (e.g., \citealp{Haardt:1991ys}) surrounding the supermassive black hole (SMBH). The X-ray spectra of AGN are usually characterised by a continuum power law, with a photon index of $\Gamma \sim 1.8-2$ (e.g., \citealp{Dadina:2008uq}, \citealp{Beckmann:2009fk}, \citealp{Corral:2011uq}), a soft excess at $\lesssim 1$\,keV (e.g., \citealp{Turner:1989zr}) and reflection features arising from circumnuclear material (e.g., \citealp{Matt:1991ly}). According to the classical unification scheme of AGN \citep{Antonucci:1993kb} the SMBH and the accretion disk are surrounded by a molecular torus which provides anisotropic obscuration, so that Seyfert\,1s (Sy1s, or type-I AGN) are observed pole-on and usually unabsorbed in the X-rays, while Seyfert\,2s (Sy2s, or type-II AGN) are seen edge-on, and usually present signatures of absorption in the X-rays. Reflection of the X-ray emission from neutral material creates a Compton hump at $\sim 30$\,keV and a fluorescent Fe\,K$\alpha$ line (e.g., \citealp{Lightman:1988ve}). Reprocessed emission is a powerful tool to understand the geometry of the material surrounding the SMBH, and is of fundamental importance to explain the cosmic X-ray background (CXB, \citealp{Giacconi:1962cr}). The spectrum of the CXB peaks at $\sim 30$\,keV (e.g., \citealp{Marshall:1980ys}, \citealp{Turler:2010fk}), and the existence of a large fraction ($\sim 30\%$) of Compton-thick (CT, $N_{\rm\,H}>10^{24}\rm\,cm^{-2}$) AGN has been invoked to explain its shape (e.g., \citealp{Gilli:2007qf}). This fraction is however strictly linked to the amount of reprocessed radiation assumed (\citealp{Treister:2009uq}), so that larger values of the reflection parameter would imply a lower fraction of CT AGN (\citealp{Gandhi:2007fk}, \citealp{Ricci:2011zr}, \citealp{Vasudevan:2013ys}). 

The narrow component of the Fe\,K$\alpha$ line peaks at 6.4\,keV (e.g., \citealp{Yaqoob:2004qf}), and it has been found to be omnipresent in the X-ray spectra of local AGN (e.g., \citealp{Shu:2012fk}). Given its full width half maximum (FWHM) of $\simeq 2000\rm\,km\,s^{-1}$ \citep{Shu:2011fk} this narrow component has been often associated to material located in the molecular torus (see also \citealp{Ponti:2013vn}). The narrow iron K$\alpha$ line has been found ubiquitously also in several high-redshift samples through spectral stacking: {\it Chandra} Deep Field North and South ($0.5<z<4$, \citealp{Brusa:2005qf}), {\it XMM-Newton} medium survey ($z<3.5$, \citealp{Corral:2008kx}), {\it XMM-Newton} bright survey ($z<2.4$, \citealp{Corral:2011uq}), 2XMM catalog ($z<5$, \citealp{Chaudhary:2012ys}), COSMOS ($z<4$, \citealp{Iwasawa:2012zr}), and {\it XMM-Newton} observations of the {\it Chandra} Deep Field South ($z<3.5$, \citealp{Falocco:2013ly}).

The aim of this paper is to study, using the results of different works performed with {\it Suzaku} \citep{Fukazawa:2011fk}, {\it XMM-Newton} \citep{Ricci:2014vs} and {\it Chandra} \citep{Shu:2010zr} data, the characteristics of the narrow Fe\,K$\alpha$ line in type-I and type-II AGN. The paper is structured as follows. In Sect.\,\ref{Sect:luminosities} we study the differences in the average ratio of the Fe\,K$\alpha$ and the X-ray continuum flux between type-I and Compton-thin type-II AGN, discuss the fact that contamination from the broad Fe K$\alpha$ line is likely to be negligible in our work, and show that it can be explained by means of different average inclination angles, as foreseen by the unified model. In Sect.\,\ref{Sect:absreflection} we show that part of the Fe\,K$\alpha$ flux is depleted in CT AGN, and estimate the average column density that absorbs the reflected component; in Sect.\,\ref{Sect:XBE} we report for the first time the existence of a X-ray Baldwin effect in type-II AGN, with the same characteristics as those observed in type-I AGN. Finally, in Sect.\,\ref{Sect:summary} we present our conclusions. Throughout the paper we refer to Compton-thin type-II AGN as Sy2s (or type\,II AGN), and to CT objects as CT Sy2s.

\section{The Fe\,K$\alpha$ line and the X-ray continuum emission }\label{Sect:luminosities}

\subsection{Seyfert\,1s}
The iron K$\alpha$ line is created by reprocessing of the primary X-ray continuum, so that a tight correlation between the luminosity of the line and that of the continuum is expected. Fitting the data of the Sy1/1.5 {\it XMM-Newton} sample of \citet{Ricci:2014vs} with a relation of the type
\begin{equation}\label{eq:Lumirvscont}
\log L_{\rm\,K\alpha} = \alpha+\beta\log L_{\,2-10},
\end{equation}
we obtained a slope of $\beta=0.89\pm0.04$. As a comparison we fitted with Eq.\,\ref{eq:Lumirvscont} the {\it Chandra}/HEG data reported in \citet{Shu:2010zr}, and obtained a consistent slope of $\beta=0.86\pm0.01$. In the left panel of Fig.\,\ref{Fig:LcLkalpha} the two data sets with their respective fits are shown. The values of the Fe\,K$\alpha$ EW reported in the {\it Chandra}/HEG study of \citet{Shu:2010zr} are significantly lower than those obtained with {\it XMM-Newton}. This might be related to the higher energy resolution of {\it Chandra}/HEG, which makes it better suited than {\it XMM-Newton}/EPIC to resolve the narrow core of the line. The Fe\,K$\alpha$ line detected by CCD-operating instruments like EPIC might, at least in some cases, have contributions from material located closer to the SMBH than the molecular torus, i.e. in the BLR or in the outer part of the accretion disk. Another possible source of flux is the Compton shoulder, a feature expected if the material where the line originates is Compton-thick (e.g., \citealp{Matt:2002vn}, \citealp{Yaqoob:2011fk}).  The difference in the Fe\,K$\alpha$ EW measured with different instruments is also evident looking at the normalisation of the X-ray Baldwin effect reported in Table\,1 of \citet{Ricci:2013fk}: the value obtained by the {\it XMM-Newton}/EPIC sample of \citet{Bianchi:2007vn} is significantly larger than that reported by the {\it Chandra}/HEG study of \cite{Shu:2010zr}.

\begin{figure*}
\centering
\begin{minipage}[!b]{.48\textwidth}
\centering
\includegraphics[width=9cm]{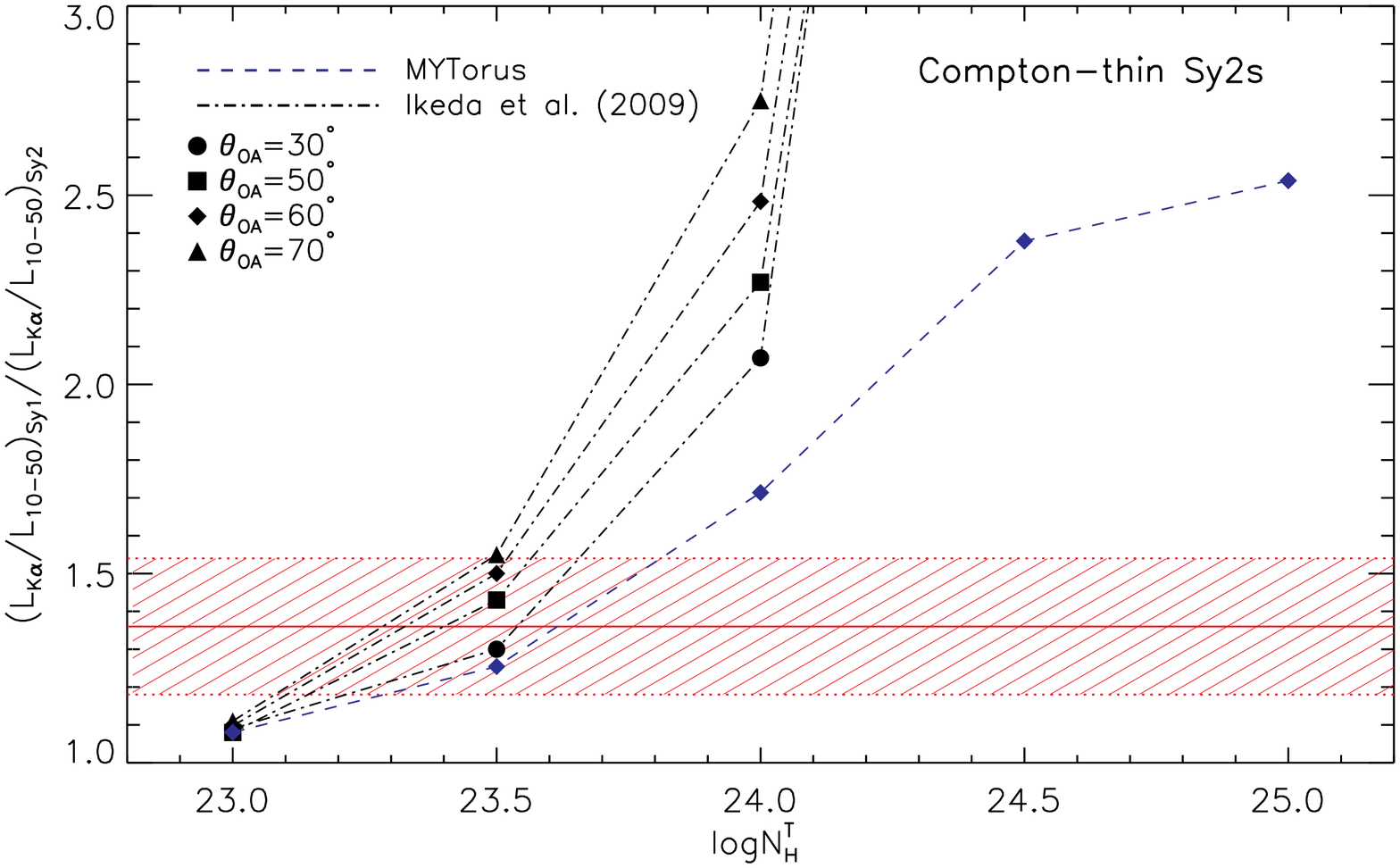}
\end{minipage}
\hspace{0.05cm}
\begin{minipage}[!b]{.48\textwidth}
\centering
\includegraphics[width=9cm]{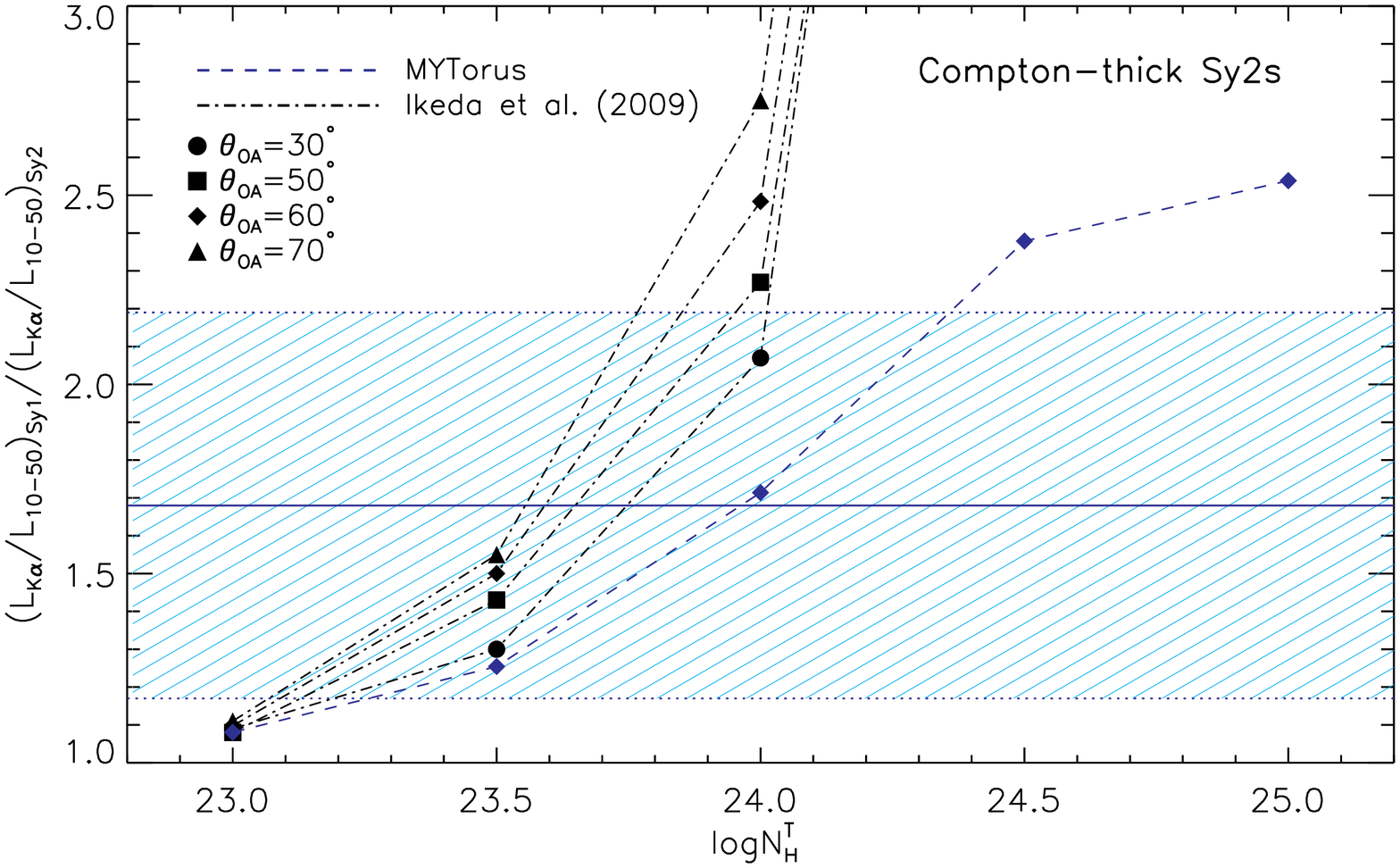}
\end{minipage}
 \begin{minipage}[t]{1\textwidth}
 \caption{{\it Left panel:} average values of the ratio between the $L_{\rm\,K\alpha}/L_{10-50}$ of Sy1s and Sy2s expected for different values of the equatorial column density of the torus ($N_{\rm\,H}^{\rm\,T}$) and of the opening angle of the torus ($\theta_{\rm\,OA}$). The blue dashed line represents the average values obtained simulating Sy1s and Sy2s populations using \texttt{MYTorus}, while the black dash-dotted line represents those obtained using the model of \citet{Ikeda:2009nx}. The red continuous line represents the average value of $(L_{\rm\,K\alpha}/L_{10-50})_{\rm\,Sy1}/(L_{\rm\,K\alpha}/L_{10-50})_{\rm\,Sy2}$ for the Compton-thin Sy2s of the {\it Suzaku} sample of \citet{Fukazawa:2011fk}, while the red shaded area is its 1$\sigma$ error. {\it Right panel}: same as left panel, with the blue continuous line and the blue shaded area representing the average value of $(L_{\rm\,K\alpha}/L_{10-50})_{\rm\,Sy1}/(L_{\rm\,K\alpha}/L_{10-50})_{\rm\,Sy2}$ for CT AGN and its 1$\sigma$ error, respectively.} \label{Fig:expected_ratio}
 \end{minipage}
\end{figure*}

\subsection{Seyfert\,1s vs Seyfert\,2s}\label{Sect:Sy1vsSy2}
In a recent work \citet{Liu:2010fk}, using [OIV] luminosity ($L_{\rm\,[OIV]}$) as a proxy for the bolometric output of AGN, found evidence of a significant difference between the Fe\,K$\alpha$ luminosities of Seyfert\,1s and Seyfert\,2s AGN. In their work, studying {\it XMM-Newton} spectra, they obtained that the values of $L_{\rm\,K\alpha}$ in Compton-thin and Compton-thick Seyfert\,2s are $2.9^{+0.8}_{-0.6}$ and $5.6^{+1.9}_{-1.4}$ times weaker than those of Seyfert\,1s. \citet{Liu:2010fk} argued that such a difference might be due to the anisotropic emission of the Fe\,K$\alpha$ line, consistently with what is expected with a line produced in the molecular torus and with the idea that Seyfert\,2s are observed edge-on with respect to the torus. However, \citet{Shu:2011fk} using the same approach and better quality {\it Chandra}/HEG data found only marginal differences between Seyfert\,1s and Seyfert\,2s. A problem with this approach is that the relation between $L_{\rm\,[OIV]}$ and the X-ray luminosity has a large dispersion \citep{Diamond-Stanic:2009oq}, which is bound to introduce a significant scatter in the observed [OIV]/Fe\,K$\alpha$ luminosity trend.

An alternative approach to study the behaviour the Fe\,K$\alpha$ luminosity in different types of AGN is to compare it to the hard X-ray ($> 10\rm\,keV$) continuum luminosity. In the hard X-ray band photons are in fact less affected by absorption than at lower X-ray energies. In order to do so we used the results of \citet{Fukazawa:2011fk}, who carried out a large study of Fe\,K$\alpha$ lines using {\it Suzaku} \citep{Mitsuda:2007kx}. Of the 87\footnote{In their work \citet{Fukazawa:2011fk} report 88 AGN, but NGC\,1142 and Swift\,J0255.2$-$0011 are the same source.} AGN reported in the work of \citet{Fukazawa:2011fk} we excluded the seven objects for which the Fe\,K$\alpha$ line was not detected and one object (NGC\,4968) for which the column density $N_{\rm\,H}$ was not constrained. The final sample consists of 30 Seyfert\,1s, 34 Compton-thin Seyfert\,2s, 13 Compton-thick Seyfert\,2s and 2 LINERs. We averaged the values of $L_{\rm\,K\alpha}$, $N_{\rm\,H}$, and of the 10--50\,keV luminosity of the continuum ($L_{10-50}^{\rm\,obs}$) for the 11 objects for which more than one observation of the same source was available. The fluxes of the X-ray continuum reported by \citet{Fukazawa:2011fk} are not corrected for absorption. Although in the 10--50\,keV band the influence of photo-electric absorption is weaker than at lower energies, Compton scattering still plays an important role, in particular for Compton-thick sources. Assuming a power-law continuum with $\Gamma=1.9$, in the 10--50\,keV band the observed flux is 91\%, 75\%, 41\%, and 7\% of the intrinsic value for column densities of $\log N_{\rm\,H}=23,23.5,24$ and 24.5, respectively.

Correcting the luminosity for absorption requires assumptions on the fraction of unabsorbed radiation reflected from material located outside the line of sight. To include self-consistently both photo-electric absorption and Compton scattering, we used \texttt{MYTorus} \citep{Murphy:2009uq} to calculate the corrections for Seyfert\,2s and CT Sy2s, similarly to what was done by \citet{Burlon:2011uq}. \texttt{MYTorus} assumes a toroidal geometry with the half-opening angle $\theta_{\rm\,OA}$ fixed to $60^{\circ}$, and its main parameters are the photon index $\Gamma$, the inclination angle of the observer $\theta_{\rm\,i}$ and the equatorial column density of the torus $N_{\rm\,H}^{\mathrm{\,T}}$. For each source, the corrections $k(N_{\rm\,H})$ were obtained from the ratio between the flux in the 10--50\,keV band of an unabsorbed ($\theta _{\rm\,i}=30^{\circ}$) and an absorbed ($\theta _{\rm\,i}=90^{\circ}$) spectrum simulated using \texttt{MYTorus}, with $N_{\rm\,H}^{\rm\,T}=N_{\rm\,H}$. The unabsorbed luminosities were then calculated multiplying the observed values by the correction ($L_{10-50}=k(N_{\rm\,H})\times L_{10-50}^{\rm\,obs}$). In the following we will only use the absorption-corrected 10--50\,keV luminosity ($L_{10-50}$). The values of $L_{\rm\,K\alpha}$ and $L_{10-50}$ used here are reported in Appendix\,\ref{list_sources}.

Fitting separately the Seyfert\,1s and Compton-thin Seyfert\,2s {\it Suzaku} samples with
 \begin{equation}\label{eq:Lumirvscont2}
\log L_{\rm\,K\alpha} = \alpha_{\mathrm{\,H}}+\beta_{\mathrm{\,H}}\log L_{\,10-50},
\end{equation}

we obtained $\alpha_{\rm\,H,1}=2.6\pm0.3$ and $\alpha_{\rm\,H,2}=2.3\pm0.5$, and identical slopes $\beta_{\mathrm{\,H,1}}=\beta_{\mathrm{\,H,2}}=0.89\pm0.01$. The fact that $\beta < 1$ implies the existence of an anti-correlation between $L_{\rm\,K\alpha}/L_{\,10-50}$ and $L_{\,10-50}$ (i.e., the X-ray Baldwin effect, see Sect.\,\ref{sect:explainingXBE}). In the right panel of Fig.\,\ref{Fig:LcLkalpha}, the scatter plot of $L_{\rm\,K\alpha}$ versus $L_{10-50}$, together with the fit for the Seyfert\,1s and Seyfert\,2s subsamples, is shown. 

As it can be seen from Fig.\,\ref{Fig:LcLkalpha}, the average Fe\,K$\alpha$ luminosity normalised to the continuum luminosity of Sy2s is lower than that of Sy1s. The mean values of $L_{\rm\,K\alpha}/L_{\,10-50}$ are in fact significantly different: $0.0047\pm0.0004$ and $0.0034\pm0.0004$ for Sy1s and Sy2s, respectively. Performing a Kolmogorov-Smirnov test we obtained a probability of $0.1\%$ that the two distributions of $L_{\rm\,K\alpha}/L_{\,10-50}$ are drawn from the same parent population.
In the following we discuss several possible explanations of this effect.

\begin{figure}
\centering
\centering
\includegraphics[width=9cm]{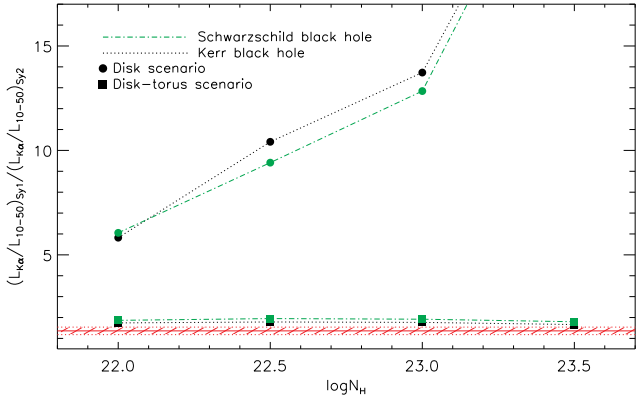}
 \caption{The points represent the average values of the ratio between the $L_{\rm\,K\alpha}/L_{10-50}$ of Sy1s and Sy2s expected for different values of the observed line-of-sight column density of Sy2s, assuming the Fe K$\alpha$ line created entirely by reprocessing of the X-ray emission in the accretion disk (circles, $R_{\rm\,disk}=1$), or half in the accretion disk and half in the molecular torus (squares, $R_{\rm\,torus}=R_{\rm\,disk}=0.5$). 
The values of $L_{\rm\,K\alpha}/L_{10-50}$ were calculated for a Schwarzschild (green dot-dashed line) and a Kerr black hole (dotted black line) (see Sect.\,\ref{sect:broadlines}), assuming that the inner radius of the disk corresponds to the radius of the inner stable circular orbit. The red continuous line represents the average value of $(L_{\rm\,K\alpha}/L_{10-50})_{\rm\,Sy1}/(L_{\rm\,K\alpha}/L_{10-50})_{\rm\,Sy2}$ for the Compton-thin Sy2s of the {\it Suzaku} sample of \citet{Fukazawa:2011fk}, while the red shaded area is its 1$\sigma$ error.} \label{Fig:broadlines}
\end{figure}

\subsubsection{Different inclination angles}\label{Sect:inclinationAngles}
We tested the hypothesis that a different average inclination angle with respect to an obscuring torus between Sy1s and Sy2s, as foreseen by the unified model, could explain this difference. We created dummy Sy2 and Sy1 populations using \texttt{MYTorus} ($\theta_{\rm\,OA}=60^{\circ}$) and the model of \citet{Ikeda:2009nx}, which considers a spherical-toroidal geometry with a varying half-opening angle for the reprocessing material. We adopted four different values of $\theta_{\rm\,OA}$ for the model of \citeauthor{Ikeda:2009nx} (\citeyear{Ikeda:2009nx}, $\theta_{\rm\,OA}=30^{\circ}, 50^{\circ}, 60^{\circ}, 70^{\circ}$), and assumed that the two populations have on average the same $\theta_{\rm\,OA}$. The value of $\theta_{\rm\,OA}$ is thought to depend on the luminosity of the AGN (e.g., \citealp{Ueda:2003nx}, see also Sect.\,\ref{sect:explainingXBE}). Performing a Kolmogorov-Smirnov test we obtained a probability of $89\%$ that the luminosity distributions of the type-I and type-II AGN of our sample are drawn from the same parent population, which makes our assumption reasonable. We fixed $\Gamma=1.9$ and, for different values of $N_{\rm\,H}^{\rm\,T}$, we generated a large number of spectra with $\theta_{\rm\,i}$ randomly selected between $1^{\circ}$ and $\theta_{\rm\,OA}$ for Sy1s, and between $\theta_{\rm\,OA}$ and $90^{\circ}$ for Sy2s. We calculated the weighted average ratio between the luminosity of the Fe\,K$\alpha$ line and the unabsorbed 10--50\,keV luminosity, using as weights $w= \sin \theta_{\rm\,i}$ to take into account the probability of observing an AGN with an angle of $\theta_{\rm\,i}$. In the left panel of Fig.\,\ref{Fig:expected_ratio}, we illustrate the expected values of $(L_{\rm\,K\alpha}/L_{10-50})_{\rm\,Sy1}/(L_{\rm\,K\alpha}/L_{10-50})_{\rm\,Sy2}$ together with the value obtained using the {\it Suzaku} results of \citet{Fukazawa:2011fk}. The figure shows that the difference observed between Sy1s and Sy2s can be explained by means of different average inclination angles between the two populations. The difference in the trends predicted by \texttt{MYTorus} and the model of \cite{Ikeda:2009nx} are related to the different geometries considered. In the spherical-toroidal scenario the surface illuminated by the X-ray source is larger than in the case of a toroidal geometry, and a larger fraction of radiation is reflected towards a pole-on observer than in the direction of an observer located edge-on. This is responsible for the high values of $(L_{\rm\,K\alpha}/L_{10-50})_{\rm\,Sy1}/(L_{\rm\,K\alpha}/L_{10-50})_{\rm\,Sy2}$ expected by the model of \cite{Ikeda:2009nx} for $\log N_{\rm\,H}^{\rm\,T}\geq 23.5$.

\begin{figure}
\centering
\centering
\includegraphics[width=.48\textwidth]{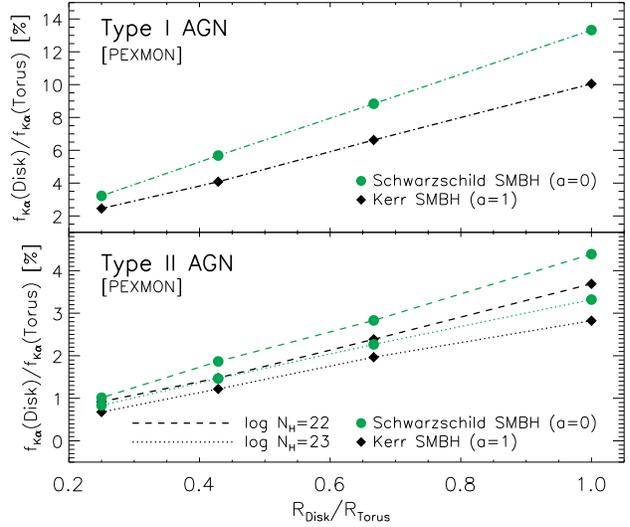}
 \caption{Ratio between the flux of the broad and narrow components of the Fe K$\alpha$ in the 6.35--6.45\,keV range expected for type\,I (top panel) and type\,II AGN (bottom panel), for different values of the ratio between the disk and torus reflection parameters ($R_{\rm\,disk}/R_{\rm\,torus}$), of the line of sight column density and of the spin of the SMBH, assuming that the inner radius of the disk corresponds to the radius of the inner stable circular orbit. The model used for the disk and torus reflection is \texttt{pexmon}. The disk reflection was blurred using the \texttt{relconv} kernel.} \label{Fig:ratio_broadnarrow}
\end{figure}

\subsubsection{Differences in Compton humps, abundances and photon index distributions}\label{Sect:Comptonhumps_ab_gam}

\citet{Ricci:2011zr}, studying the average hard X-ray (17--250\,keV) spectra of AGN, found that mildly obscured Sy2s (MOB, $23 \leq \log N_{\rm\,H}< 24$) show a stronger Compton hump than lightly obscured objects (LOB, $\log N_{\rm\,H}\leq 23$). This trend was recently confirmed by stacking 14--195\,keV {\it Swift}/BAT spectra by \citet{Vasudevan:2013ys}, and has been interpreted as being due to the presence of partially covering CT material in the line of sight, or to a larger covering factor of the torus in MOB Sy2s. The {\it Suzaku} Sy2s sample is dominated by MOB Sy2s, with only three LOB objects. Similarly to Sy1s, the three LOB AGN have an average ratio of Fe K$\alpha$ to continuum luminosity ($L_{\rm\,K\alpha}/L_{10-50}=0.0048$) larger than that of MOB Sy2s ($L_{\rm\,K\alpha}/L_{10-50}=0.0033$). A larger Compton hump in MOB Sy2s would also imply larger values of $L_{10-50}$. This might lead to a lower $L_{\rm\,K\alpha}/L_{10-50}$ ratio for MOB Sy2s, in agreement with our results. However the relation between the Fe K$\alpha$ line and the Compton hump luminosities depends on the origin of the enhanced reflection, so that we cannot estimate the impact of a stronger Compton hump on the differences in the values of $L_{\rm\,K\alpha}/L_{10-50}$. Different distributions of photon indices between Sy1s and Sy2s might also explain the different distributions of $L_{\rm\,K\alpha}/L_{10-50}$. 

Another possible explanation might be the existence of differences in the iron abundances of the circumnuclear material of different classes of objects. While it not possible to fully discard this possibility, we argue that currently there is no evidence that could confirm it.

\begin{figure*}
\centering
\begin{minipage}[!b]{.48\textwidth}
\centering
\includegraphics[width=8.8cm]{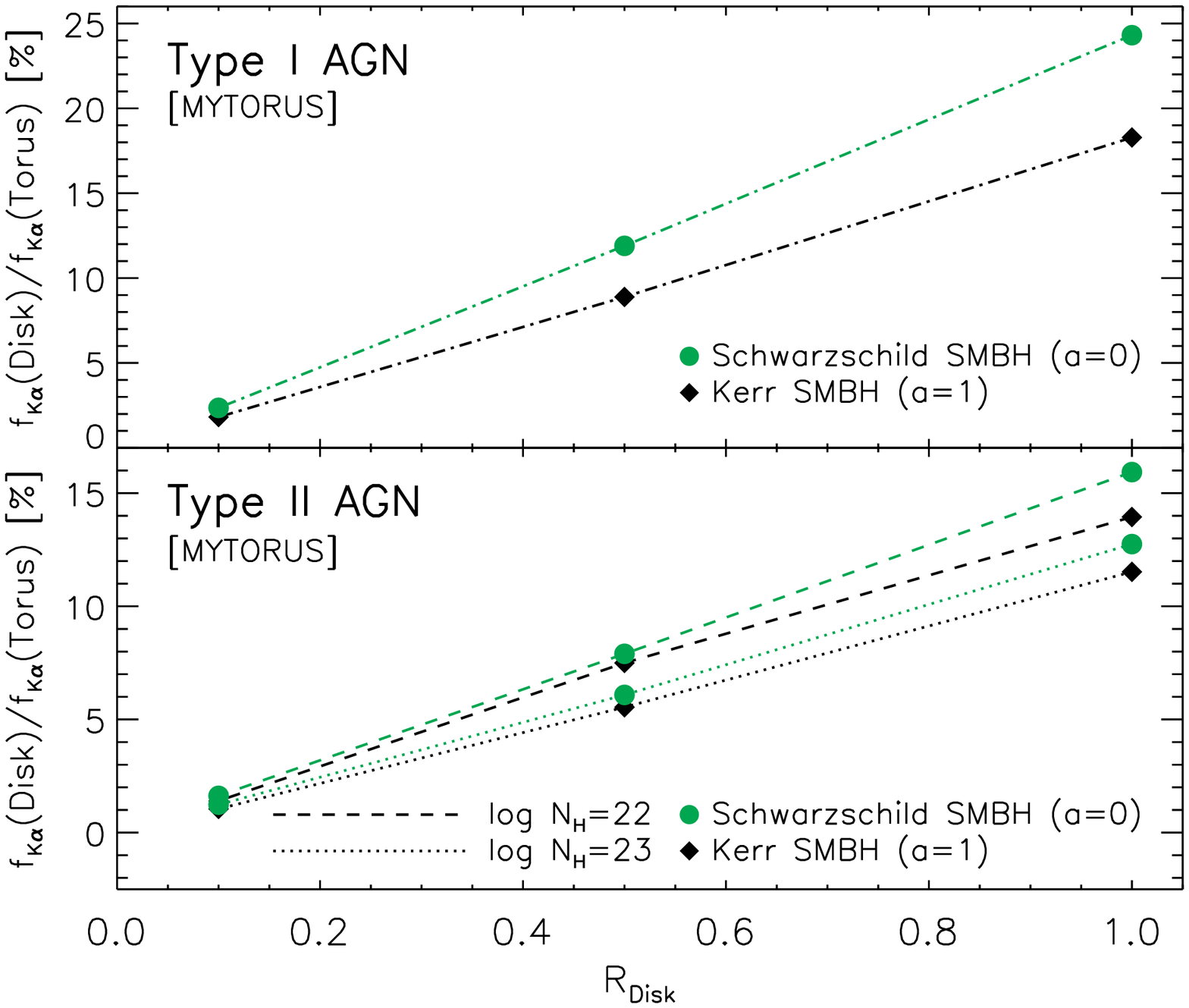}
\end{minipage}
\hspace{0.05cm}
\begin{minipage}[!b]{.48\textwidth}
\centering
\includegraphics[width=8.8cm]{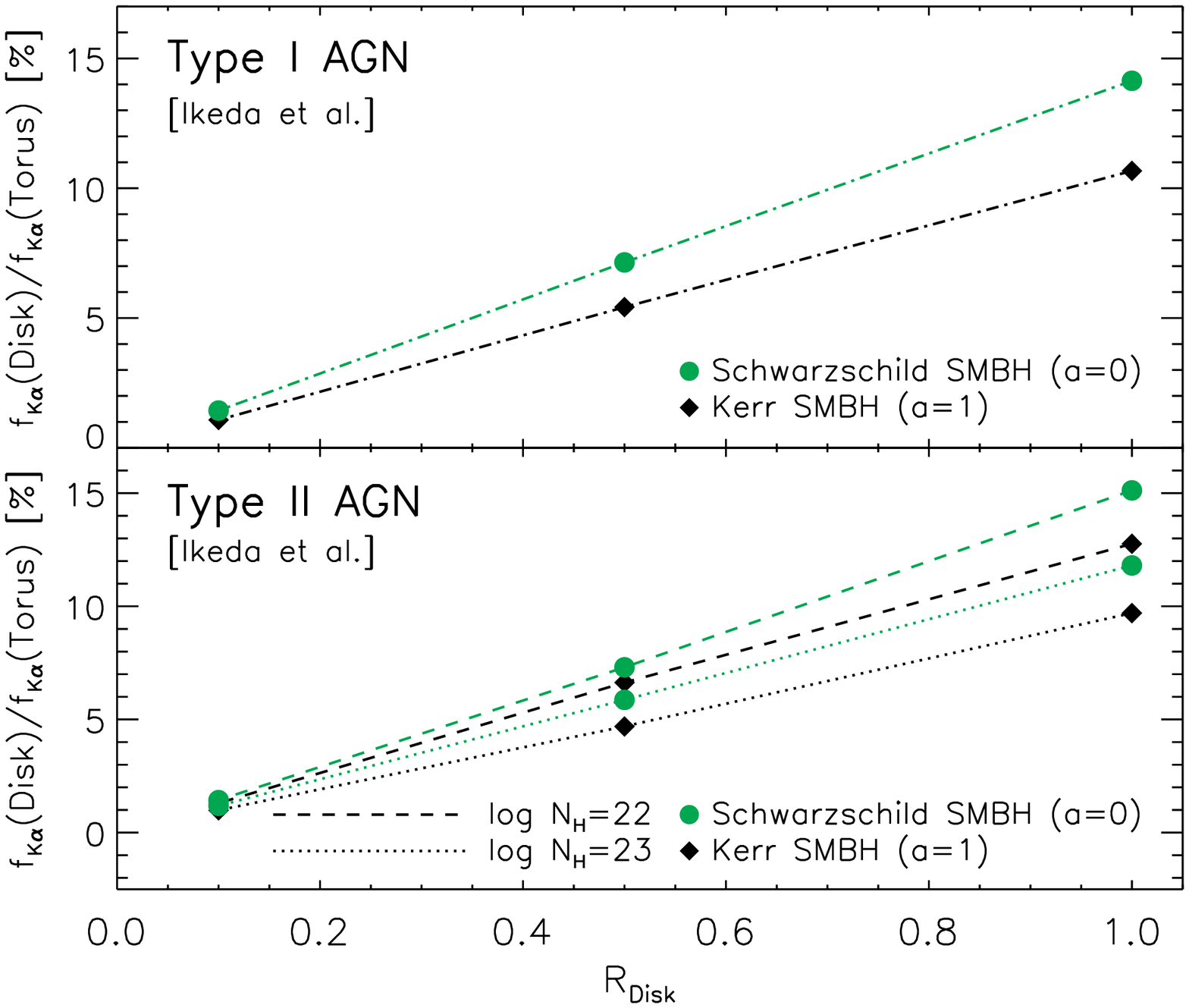}
\end{minipage}
 \begin{minipage}[t]{1\textwidth}
 \caption{{\it Left panel:} Ratio between the flux of the broad and narrow components of the Fe K$\alpha$ in the 6.35--6.45\,keV range expected for type\,I (top panel) and type\,II AGN (bottom panel), for different values of the disk reflection parameter ($R_{\rm\,disk}$), of the line of sight column density and of the spin of the SMBH, assuming that the inner radius of the disk corresponds to the radius of the inner stable circular orbit. The model used for the blurred disk reflection is \texttt{pexmon}, convolved with the \texttt{relconv} kernel. The torus reflection was simulated using \texttt{MYTORUS}, fixing $N_{\rm\,H}^{\rm\,T}=10^{24}\rm\,cm^{-2}$. {\it Right panel}: same as left panel using the model of \citet{Ikeda:2009nx} to simulate the reflection from the torus (fixing $N_{\rm\,H}^{\rm\,T}=10^{24}\rm\,cm^{-2}$ and $\theta_{\rm\,OA}=60^{\circ}$).}\label{Fig:ratio_broadnarrow_torusmod}
 \end{minipage}
\end{figure*}

\subsubsection{The role of broad lines}\label{sect:broadlines}
In the work of \cite{Fukazawa:2011fk} the width of the Fe K$\alpha$ line was not fixed. Although the lines observed were narrow ($\sigma < 80$\,eV), there might have been some contamination from the broad relativistic lines produced in the inner part of the accretion flow. Here we discuss the possibility that the differences in the $L_{\rm\,K\alpha}/L_{10-50}$ ratios between type\,I and type\,II AGN is due to the different contribution of broad lines to the observed Fe K$\alpha$ flux in type-I and type-II AGN. Both the inclination angle and the absorbing material might play a role in decreasing the intensity of the ratio between Fe K$\alpha$ line and the X-ray continuum luminosity in Sy2s. The absorbing material in the line of sight would in fact deplete the broad component of the Fe K$\alpha$ line in Sy2s, while the albedo of the disk decreases for increasing values of the inclination angle of the observer.

To test the influence of broad lines we carried out extensive simulations adopting the following approach. We simulated in XSPEC spectra in the 0.3--10\,keV range using a power-law continuum plus a blurred reprocessed emission, which includes a Fe K$\alpha$ line. The X-ray reprocessed emission was accounted for using the \texttt{pexmon} model \citep{Magdziarz:1995pi,Nandra:2007ly}, fixing the cutoff energy to $E_{\rm\,C}=300\rm\,keV$, and the value of the reflection parameter to a negative value, which allows to take into account only the reflected radiation and not the continuum. This reflected emission was then blurred using the \texttt{relconv} kernel \citep{Dauser:2010ys}, assuming a limb-darkening law \citep{Laor:1991kx}. The outer radius of the accretion disk was set to $R_{\rm\,out}=400\,r_{\rm\,G}$, where $r_{\rm\,G}=G\,M_{\rm\,BH}/c^2$ is the gravitational radius. The inclination angle with respect to the accretion disk ($\theta_{\rm\,disk}$) was randomly selected for each simulation in the $1-60^{\circ}$ and  $60-85^{\circ}$ ranges for Sy1s and Sy2s, respectively. The choice of $85^{\circ}$ as an upper limit is related to the limitations of \texttt{pexmon}.
We tested two scenarios, one in which the spin of the SMBH is $a=0.998$ and the inner radius of the disk is $R_{\rm\,in}=1.24\,r_{\rm\,G}$, equivalent to a maximally-rotating Kerr black hole, and one in which $a=0$ and $R_{\rm\,in}=6\,r_{\rm\,G}$, as expected for a non-rotating Schwarzschild black hole. We considered four different values of the line-of-sight column density for Sy2s ($\log N_{\rm\,H}=22,22.5,23$ and $23.5$) for type\,II AGN, and took into account both photoelectric absorption (\texttt{tbabs} in XSPEC, \citealp{Wilms:2000vn}) and Compton scattering (\texttt{cabs}). We fixed the internal ($r \leq R_{\rm\,break}$) and external ($r>R_{\rm\,break}$) emissivity indices to $q_1=q_2=3$, and set the value of $R_{\rm\,break}$ to an arbitrary value.
The simulated spectra were fitted using a power-law model plus a Gaussian line, with a energy fixed to 6.4\,keV and a width that was left free to vary (with an upper bound of 300\,eV). We calculated the weighted average ratio between the luminosity of the Fe K$\alpha$ line and the unabsorbed 10--50\,keV luminosity, using the same weights adopted in Sect.\,\ref{Sect:inclinationAngles} ($w= \sin \theta_{\rm\,i}$).

In Fig.\,\ref{Fig:broadlines} we illustrate the values of $(L_{\rm\,K\alpha}/L_{10-50})_{\rm\,Sy1}/(L_{\rm\,K\alpha}/L_{10-50})_{\rm\,Sy2}$ expected for different values of $a$ and $N_{\rm\,H}$. 
We considered two cases. One in which all the emission in the Fe K$\alpha$ band is due to blurred reflection from the disk (\emph{disk scenario}, with a reflection parameter of $R_{\rm\,disk}=1$), and one in which half of the emission is created in the disk and half in the molecular torus (\emph{disk-torus scenario}). In the latter case we added a second unblurred \texttt{pexmon} to account for the reflection from neutral distant material ($R_{\rm\,torus}=R_{\rm\,disk}=0.5$), fixing the inclination angle to $\theta_{\rm\,tor}=\pi/2-\theta_{\rm\,disk}$. This second reflection component was considered unobscured when simulating the X-ray spectra of type-II AGN.
The figure shows that differences in the broad component of the Fe K$\alpha$ line would produce a higher ratio than the one observed, both in the disk and in the disk-torus scenario, which excludes a strong influence of the broad lines in our work. 
If the fluxes of the Fe K$\alpha$ lines were heavily contaminated by broad lines (disk scenario), one would expect a steep increase of the ratio $(L_{\rm\,K\alpha}/L_{10-50})_{\rm\,Sy1}/(L_{\rm\,K\alpha}/L_{10-50})_{\rm\,Sy2}$ with the line-of-sight column density for $\log N_{\rm\,H}\geq 23$. This is not observed (see Fig.\,\ref{Fig:ratiovsNH} and Sect.\,\ref{Sect:absreflection}). A scenario in which reflection from the disk and from the torus have the same influence could also be discarded, as it would lead to values of $(L_{\rm\,K\alpha}/L_{10-50})_{\rm\,Sy1}/(L_{\rm\,K\alpha}/L_{10-50})_{\rm\,Sy2}$ higher than the observed ones. This implies that on average $R_{\rm\,disk}/R_{\rm\,torus}<1$.

To estimate the ratio between the flux of broad and narrow Fe K$\alpha$ lines at 6.4\,keV we simulated two sets of dummy AGN populations, taking into account only reflection and considering $R_{\rm\,disk}/R_{\rm\,torus}<1$. One set of spectra were simulated using blurred reflection (\texttt{relconv(pexmon)} in XSPEC), while the other considering unblurred reflection (\texttt{pexmon}). We considered two different scenarios to study the ratio in type\,I and type\,II AGN. 
For type\,I AGN we randomly selected inclination angles in the $1-60^{\circ}$ and $30-85^{\circ}$ ranges for the blurred and unblurred reflection component, respectively. We then obtained the fluxes for each simulated spectrum in the 6.35--6.45\,keV range, and calculated the weighted averages for the blurred and unblurred reflection case, using the same weights adopted in Sect.\,\ref{Sect:inclinationAngles}. 
In Fig.\,\ref{Fig:ratio_broadnarrow} (top panel) we illustrate the values of the ratio between the 6.35--6.45\,keV fluxes of broad and narrow lines ($f_{\rm\,K\alpha}(\mathrm{disk})/f_{\rm\,K\alpha}(\mathrm{torus})$) for different values of $R_{\rm\,disk}/R_{\rm\,torus}$, and considering a disk accreting onto a Schwarzschild or a maximally rotating Kerr black hole. 
For type\,II AGN we followed the same approach used for type\,I, including absorption in the case of the blurred reflection\footnote{\texttt{tbabs$\times$cabs$\times$[relconv(pexmon)]}}, and selecting random angles in the $60-85^{\circ}$ and $5-30^{\circ}$ ranges for the blurred and unblurred reflection component, respectively.
As shown in Fig.\,\ref{Fig:ratio_broadnarrow}, for type\,I AGN we obtained $f_{\rm\,K\alpha}(\mathrm{disk})/f_{\rm\,K\alpha}(\mathrm{torus})<14\%$, while for type\,II AGN the influence of the broad line is even lower 
($f_{\rm\,K\alpha}(\mathrm{disk})/f_{\rm\,K\alpha}(\mathrm{torus})<5\%$). 
The values of the ratios obtained using \texttt{MYTORUS} and the model of \citet{Ikeda:2009nx} instead of \texttt{pexmon} to simulate the reflection from the torus are shown in Fig.\,\ref{Fig:ratio_broadnarrow_torusmod}. In both models the equatorial column density of the torus was fixed to $N_{\rm\,H}^{\rm\,T}=10^{24}\rm\,cm^{-2}$, and in the model of \citet{Ikeda:2009nx} the half-opening angle of the torus was set to $\theta_{\rm\,OA}=60^{\circ}$. The values of the $f_{\rm\,K\alpha}(\mathrm{disk})/f_{\rm\,K\alpha}(\mathrm{torus})$ ratios are $<25\%$ for type\,I AGN, and $<15\%$ for type\,II AGN\footnote{The relative influence of the disc line would however increase if the disc is truncated or in the presence of light bending (e.g, \citealp{Miniutti:2004ys})}. These simulations show that the influence of disk lines is on average marginal in our work.

Further evidence that the broad Fe K$\alpha$ line does not play a significant role in our work comes from the fact that the width of the lines does not decrease with increasing values of $N_{\rm\,H}$, and the mean width of the Fe K$\alpha$ line obtained for different AGN classes are consistent: $\sigma_{\,Sy1}=79$\,eV and $\sigma_{\,Sy2}=64$\,eV, with a standard deviation of 37 and 33\,eV, respectively. Our simulations show that significant differences in the width of the Fe K$\alpha$ line between type\,I and type\,II AGN are expected if the disk reflection is as strong as the torus reflection. The idea that the contribution of broad lines to our work is marginal is also supported by the fact that the values of $\sigma$ are much smaller than those obtained by \cite{Patrick:2012zr} ($\sigma\simeq 500$\,eV) studying relativistic disk lines using {\it Suzaku}, and by the fact that the energy of the Gaussian lines reported by \cite{Fukazawa:2011fk} is on average $E= 6.399$\,keV, a value in good agreement with neutral or low-ionisation Fe K$\alpha$ fluorescent emission.

\section{Absorption of reprocessed emission}\label{Sect:absreflection}

In the $\log L_{\rm\,K\alpha}-\log L_{\,10-50}$ plane, CT Sy2s have a significantly flatter slope ($\beta_{\mathrm{\,H,CT}}=0.36\pm0.02$) than Sy1s and Sy2s, and as shown in Fig.\,\ref{Fig:LcLkalpha} (right panel) their Fe\,K$\alpha$ luminosity is systematically lower than that observed in Sy1s and Sy2s for the same continuum luminosity. 
We tested the idea that this difference could be explained in terms of different inclination angles. Comparing the ratios of the Fe K$\alpha$ and the continuum luminosities of CT Sy2s and Sy1s we obtained an average value of $(L_{\rm\,K\alpha}/L_{10-50})_{\rm\,Sy1}/(L_{\rm\,K\alpha}/L_{10-50})_{\rm\,CT}=1.68\pm 0.51$, consistent with the theoretical predictions for a toroidal structure of the reprocessing and absorbing material, but not for a spherical-toroidal geometry (see right panel of Fig.\,\ref{Fig:expected_ratio}), because $N_{\rm\,H}^{\rm\,T}$ cannot be lower than $N_{\rm\,H}$. 

\begin{figure}
\centering
\centering
\includegraphics[width=9cm]{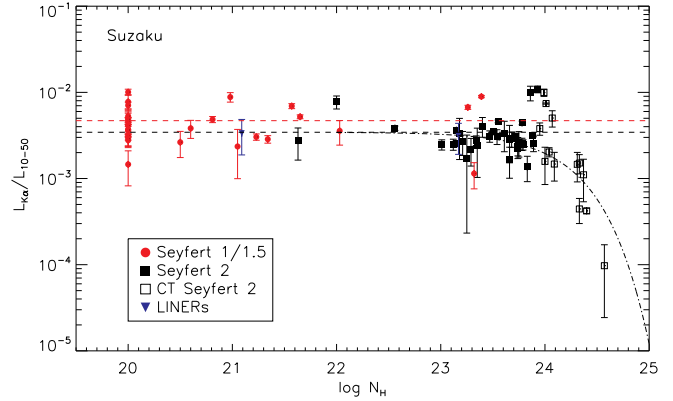}
 \caption{Ratio between the Fe\,K$\alpha$ and the continuum luminosity ($L_{\rm\,K\alpha}/L_{10-50}$) versus the observed hydrogen column density ($N_{\rm\,H}$) for the {\it Suzaku} sample of \citet{Fukazawa:2011fk}. The continuum luminosities were corrected to account for photoelectric absorption and Compton scattering as described in Sect.\,\ref{Sect:Sy1vsSy2}. The unabsorbed objects were arbitrarily assigned $\log N_{\rm\,H}=20$. The red and black dashed lines represent the average of the ratio $L_{\rm\,K\alpha}/L_{10-50}$ for Seyfert\,1s and Seyfert\,2s, respectively. The dot-dashed black line represents the best fit to the Sy2s and CT Sy2s data obtained using Eq.\,\ref{eq:ratiovsNHctExp}.} \label{Fig:ratiovsNH}
\end{figure}

An alternative explanation is that in these objects part of the reflected radiation is absorbed. Plotting the $L_{\rm\,K\alpha}/L_{10-50}$ ratio versus the observed column density (Fig.\,\ref{Fig:ratiovsNH}), we found a steep decrease of the relative iron K$\alpha$ emission above $N_{\rm\,H}\sim 10^{24}\rm\,cm^{-2}$. Fitting this decrease (for $\log N_{\rm\,H}\geq 24$) with a log-linear relation we obtained
\begin{equation}\label{eq:ratiovsNHct}
\log \frac{L_{\rm\,K\alpha}}{L_{\,10-50}} = (-3.12\pm0.08)\log N_{\rm\,H,\,22}+(4.1\pm0.2),
\end{equation}
where $N_{\rm\,H,\,22}$ is the column density in units of $10^{\,22}\rm\,cm^{-2}$. No significant trend is found in Seyfert\,2s (with a null hypothesis probability of $P_{\rm\,n}\simeq 99\%)$. The decrease of $L_{\rm\,K\alpha}/L_{10-50}$ with the observed column density in CT Sy2s is likely to be due to absorption of the reflected emission from the material located between the observer and the X-ray source, often associated to the putative molecular torus. A similar decrease is also found using the time-averaged absorption-corrected 14--195\,keV {\it Swift}/BAT luminosities (\citealp{Baumgartner:2010uq}, see Sect.\,\ref{Sect:bat}) instead of the 10--50\,keV {\it Suzaku} luminosities.

From the decrease of $L_{\rm\,K\alpha}/L_{10-50}$ with $N_{\rm\,H}$ it is possible to obtain the average fraction of the line-of-sight column density that absorbs the X-ray radiation reprocessed in the molecular torus ($X_{\rm\,R}$).  This can be done by fitting the data of Sy2s and CT Sy2s with the following relation:
\begin{equation}\label{eq:ratiovsNHctExp}
 \frac{L_{\rm\,K\alpha}}{L_{\,10-50}}= \frac{L_{\rm\,K\alpha}^{\rm\,unabs}}{L_{\,10-50}}\times \exp{\left\{-[\sigma_{\rm\,T}+\sigma_{\rm\,ph}(6.4\rm\,keV)]\times X_{\rm\,R}\times N_{\rm\,H}\right\}},
\end{equation}
where $L_{\rm\,K\alpha}^{\rm\,unabs}$ is the unabsorbed luminosity of the Fe K$\alpha$ line, $\sigma_{\rm\,T}$ is the Thomson cross section, and $\sigma_{\rm\,ph}(6.4\rm\,keV)$ is the photoelectric cross section at 6.4\,keV. The value of $\sigma_{\rm\,ph}$ at 6.4\,keV was calculated using the coefficients of the analytic fit in the 4.038-7.111\,keV energy band reported by \citet{Morrison:1983fk}. We fixed $L_{\rm\,K\alpha}^{\rm\,unabs}/L_{\,10-50}$ to the ratio obtained for Sy2s, as this value is constant up to $\log N_{\rm\,H}\sim 24$. From the fit we obtained $X_{\rm\,R}=0.24\pm0.01$, which implies that on average the reflected emission is absorbed by a column density which is about $1/4$ of the one that absorbs the X-ray primary emission.

\begin{table*}
\begin{center}
\caption{Intercept ($\alpha$), slope ($\beta$), and Spearman's rank coefficient ($\rho$) obtained by fitting $L_{\rm\,K\alpha}$ and the MIR luminosities of the {\it Suzaku} sample of \citet{Fukazawa:2011fk} with Eq.\,\ref{eq:LumirvsMIR} for the different subsamples. The MIR luminosities in the 3.4\,$\mu$m (1, 2, 3), 4.6\,$\mu$m (4, 5, 6), 12\,$\mu$m (7, 8, 9), and 22\,$\mu$m (10, 11, 12) band were taken from the {\it WISE} all-sky survey.}
\resizebox{\textwidth}{!}{
\label{tab:fit_MIR}
\begin{tabular}[c]{lccccccccccccccc}
\hline \hline  \\
\multicolumn{1}{l}{ Sample} & 
\multicolumn{3}{c}{ 3.4\,$\mu$m} &  \phantom{} &
\multicolumn{3}{c}{ 4.6\,$\mu$m  } &  \phantom{} &
\multicolumn{3}{c}{12\,$\mu$m} &  \phantom{} &
\multicolumn{3}{c}{22\,$\mu$m}   \\ 
\multicolumn{1}{l}{  } & 
\multicolumn{3}{l}{ } &  \phantom{} &
\multicolumn{3}{l}{} &  \phantom{} &
\multicolumn{3}{l}{} &  \phantom{} &
\multicolumn{3}{l}{}   \\ \cline{2-4} \cline{6-8} \cline{10-12} \cline{14-16}
\noalign{\smallskip}
 & (1) & (2) & (3) &\phantom{} &  (4) & (5) & (6) &\phantom{} & (7) & (8) & (9) &\phantom{} & (10) & (11) & (12)  \\
\noalign{\smallskip}
 & $\alpha_{\mathrm{MIR}}$ &$\beta_{\mathrm{MIR}}$ & $\rho$ & \phantom{} & $\alpha_{\mathrm{MIR}}$ &$\beta_{\mathrm{MIR}}$& $\rho$ & \phantom{} & $\alpha_{\mathrm{MIR}}$ &$ \beta_{\mathrm{MIR}}$& $\rho$ & \phantom{} & $\alpha_{\mathrm{MIR}}$ &$ \beta_{\mathrm{MIR}}$ & $\rho$ \\
\noalign{\smallskip}
\hline \noalign{\smallskip}
Seyfert\,1s &$41.04\pm0.01$& $1.05\pm0.01$  &  0.94   &\phantom{} & $41.08\pm0.01$ &$1.00\pm0.01$ & 0.94    &\phantom{} & $41.06\pm0.01$ &$1.10\pm0.01$ &  0.94   &\phantom{} & $40.93\pm0.01$ &$1.14\pm0.01$  &0.87\\
\noalign{\smallskip}
Seyfert\,2s & $41.16\pm0.01$ &$0.96\pm0.02$ &  0.80   &\phantom{} & $41.18\pm0.01$ &$0.92\pm0.02$ &  0.79   &\phantom{} & $41.09\pm0.01$ &$0.95\pm0.02$ &  0.75   &\phantom{} & $40.88\pm0.01$ &$0.82\pm0.02$ & 0.69 \\
\noalign{\smallskip}
CT Seyfert\,2s & $40.08\pm0.01$ &$0.52\pm0.03$ &  0.76   &\phantom{} & $39.99\pm0.01$ &$0.38\pm0.02$ &  0.76   &\phantom{} & $39.83\pm0.01$ &$0.24\pm0.02$ &  0.63   &\phantom{} & $39.84\pm0.01$ &$0.47\pm0.03$  & 0.75\\
\noalign{\smallskip}
\noalign{\smallskip}
Sy1s + Sy2s & $41.10\pm0.01$ &$1.00\pm0.01$ & 0.84    &\phantom{} & $41.13\pm0.01$ &$0.96\pm0.01$ &  0.85   &\phantom{} & $41.08\pm0.01$ &$1.04\pm0.01$ &  0.84   &\phantom{} & $40.91\pm0.01$ &$1.02\pm0.01$  & 0.80 \\
\noalign{\smallskip}
\hline \noalign{\smallskip}
\end{tabular}
}
\end{center}
\end{table*}%

It should be remarked that \cite{Fukazawa:2011fk} report values of the column density of $\log N_{\rm\,H}\sim 24$ for two objects, NGC\,1068 and ESO\,323$-$G032, which have been reported by other works as being reflection-dominated, with column densities of $\log N_{\rm\,H}\geq 25$ \citep{Matt:1997cr,Comastri:2010nx}. Interestingly, these two objects have a value of $L_{\rm\,K\alpha}/L_{10-50}$ significantly larger than the average value (Fig.\,\ref{Fig:ratiovsNH}), which is probably related to the under-estimation of the intrinsic luminosity due to the low value of $N_{\rm\,H}$ reported by \cite{Fukazawa:2011fk}.

\begin{figure}
\centering
\includegraphics[width=9cm]{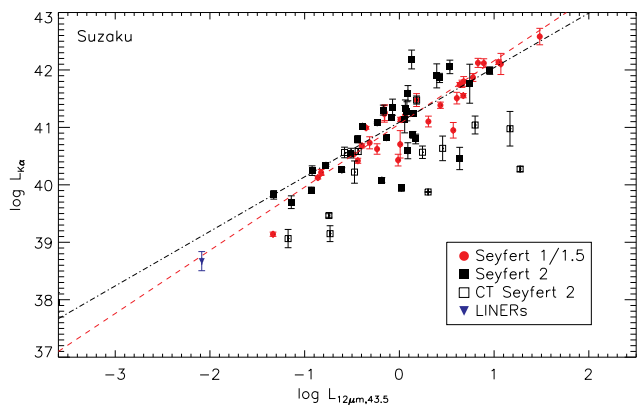}
 \caption{Fe\,K$\alpha$ luminosities versus the 12\,$\mu$m band luminosities (in units of $10^{\,43.5}\rm\,erg\,s^{-1}$) for the {\it Suzaku} sample of \citet{Fukazawa:2011fk}. The MIR luminosities were taken from the {\it WISE} all sky survey. The red dashed and the black dot-dashed lines represent the fits obtained using Eq.\,\ref{eq:LumirvsMIR} for Seyfert\,1s and Seyfert\,2s, respectively (see also Table\,\ref{tab:fit_MIR}).}
\label{Fig:lumMIR}
\end{figure}

The technique we used to correct the continuum luminosity for absorption is geometry-dependent, and takes into account also reflection from the torus. Correcting the values of the luminosity taking into account only the X-ray primary emission, and thus excluding the reflected flux, results in an even steeper decrease of $L_{\rm\,K\alpha}/L_{\rm\,10-50}$ with $N_{\rm\,H}$ for CT AGN.
An alternative approach is to compare the Fe\,K$\alpha$ luminosity of AGN to their MIR luminosities, as the emission in this band is produced by the material responsible for the flux attenuation and thus expected to be less biased by absorption (e.g., \citealp{Ichikawa:2012fk}). This was done cross-correlating the {\it Suzaku} sample with the {\it WISE} all-sky survey catalog \citep{Wright:2010fk}. Of the 79 objects of the {\it Suzaku} sample, only for the Seyfert\,2 Mrk\,3 no flux is reported in the {\it WISE} all-sky source catalog.
The MIR luminosities reported in the {\it WISE} catalog are in the 3.4\,$\mu$m, 4.6\,$\mu$m, 12\,$\mu$m, 22\,$\mu$m band.
We considered photometric values obtained by PSF-fitting, and for each of the four bands we performed a fit using
\begin{equation}\label{eq:LumirvsMIR}
\log L_{\rm\,K\alpha} = \alpha_{\mathrm{MIR}}+\beta_{\mathrm{MIR}}\log L_{\rm\,MIR,\,43.5},
\end{equation}
where $L_{\rm\,MIR,\,43.5}$ is the luminosity in units of $10^{\,43.5}\rm\,erg\,s^{-1}$.
The results of the fits are reported in Table\,\ref{tab:fit_MIR} for the different AGN subsamples. In Fig.\,\ref{Fig:lumMIR} we show the scatter plot of $L_{\rm\,K\alpha}$ versus the 12\,$\mu$m luminosity. The scatter in the correlation is larger than the one observed comparing $L_{\rm\,K\alpha}$ to $L_{\,10-50}$, but it still shows that Compton-thick sources have a lower Fe\,K$\alpha$ luminosity than Compton-thin Seyfert\,2s and Seyfert\,1s. Some of the scatter in the correlation might be related to contamination from the host galaxy, as the spatial resolution of {\it WISE} does not allow to clearly discern the emission of material located close to the central engine from that of dust on the kpc scale. Another source of scatter might be the different covering factors and equatorial column densities of the tori. Of the four bands, the 3.4\,$\mu$m and 22\,$\mu$m are the most affected by host galaxy contamination, while the 3.4\,$\mu$m and 4.6\,$\mu$m band could be subject to contamination from PAH. Using high spatial resolution photometry, \citet{Gandhi:2009uq} have shown that the intrinsic MIR emission of CT Sy2s is not significantly affected by absorption. Moreover, significant absorption would imply that the MIR luminosity of CT AGN is larger than that reported by {\it WISE}. This would move rightward the CT sources in Fig.\,\ref{Fig:lumMIR}, resulting in a Fe\,K$\alpha$ luminosity even lower than the observed one when compared to Seyfert\,1s and Seyfert\,2s. 

The fits obtained applying Eq.\,\ref{eq:LumirvsMIR} show that CT AGN have a significantly flatter slope, and that for the same MIR luminosity their Fe\,K$\alpha$ luminosities are systematically lower than those of Seyfert\,1s and Seyfert\,2s in the four bands for which {\it WISE} observations were available. This confirms the idea that in very obscured objects part of the reflected component is absorbed, and it should be taken into account when carrying out spectroscopical X-ray analysis and studies of their luminosity function. These findings should also have important consequences on synthesis models of the CXB. Absorption of part of the reflected flux would in fact imply that a larger number of CT AGN is needed. In a future work we will estimate how the fraction of CT AGN would change because of the absorption of the reprocessed radiation.

\begin{figure*}
\centering
\begin{minipage}[!b]{.48\textwidth}
\centering
\includegraphics[width=9cm]{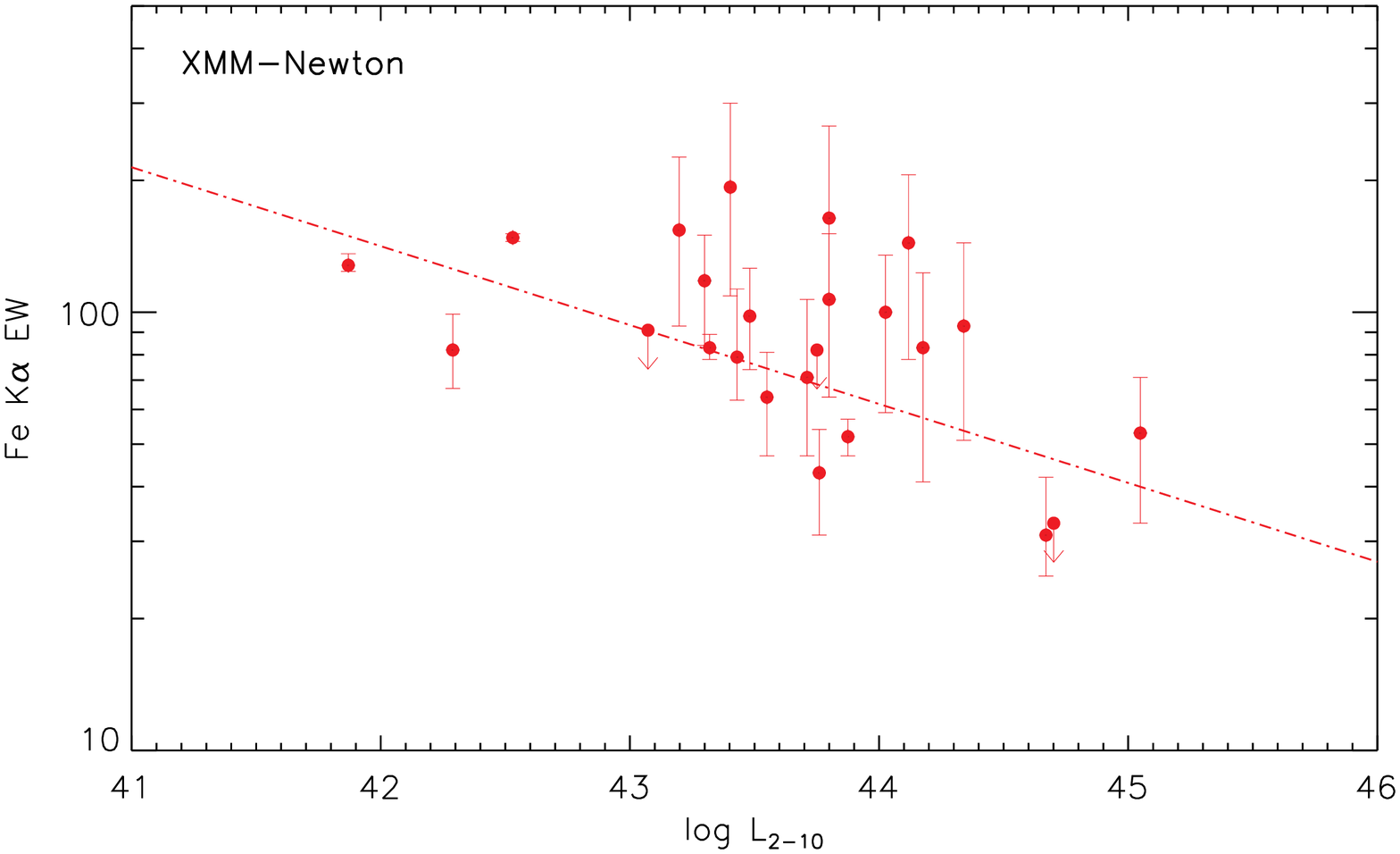}
\end{minipage}
\hspace{0.05cm}
\begin{minipage}[!b]{.48\textwidth}
\centering
\includegraphics[width=9cm]{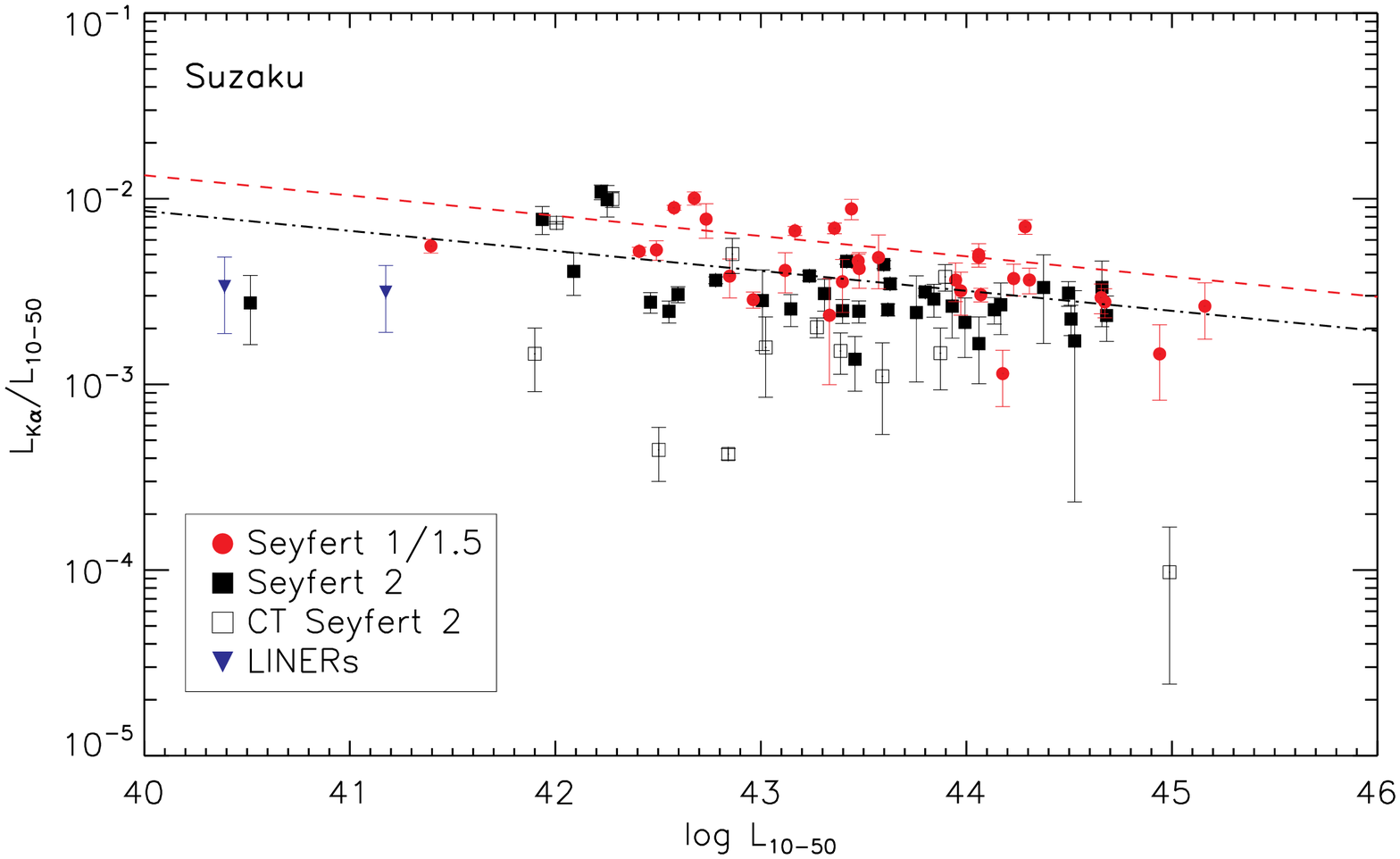}\end{minipage}
 \begin{minipage}[t]{1\textwidth}
  \caption{{\it Left panel}: Fe\,K$\alpha$ EW versus the continuum luminosity for the {\it XMM-Newton} sample of Sy1s of \citet{Ricci:2014vs}. The red dashed line represents the fit (with a slope of $\overline{\mu}= -0.16\pm 0.06$) to the data obtained using the regression method for left censoring data described in Sect.\,\ref{sect:XBEtype1type2}.  {\it Right panel}: ratio between the luminosity of the Fe\,K$\alpha$ and that of the 10--50\,keV continuum ($L_{\rm\,K\alpha}/L_{\,10-50}$) versus $L_{\,10-50}$ for the {\it Suzaku} sample of \citet{Fukazawa:2011fk}. The values of $L_{\,10-50}$ of Sy2s and Compton-thick Seyfert\,2s have been corrected for absorption as described in Sect.\,\ref{Sect:luminosities}. The red (black) dashed (dot-dashed) line represents the best fit to the data obtained using Eq.\,\ref{eq:xbe2} for Sy1s (Sy2s). The Sy1 and Sy2 sample have the same slope ($\omega_{\,1}=\omega_{\,2}=-0.11\pm0.01$). }
\label{Fig:XBE}
 \end{minipage}

\end{figure*}

\section{The X-ray Baldwin effect}\label{Sect:XBE}

\subsection{Seyfert\,1s and Seyfert\,2s}\label{sect:XBEtype1type2}
The existence of a significant anti-correlation between EW and $L_{\mathrm{\,2-10}}$ (i.e. the X-ray Baldwin effect, \citealp{Iwasawa:1993ys}) has been confirmed in the last years by several works carried out with the highest energy resolution available in the X-rays (e.g.,  \citealp{Bianchi:2007vn}, \citealp{Shu:2010zr}). 
To study the relation between Fe\,K$\alpha$ EW and $L_{\mathrm{\,2-10}}$ for the {\it XMM-Newton} sample of type-I AGN of \citet{Ricci:2014vs}, which includes several upper limits, we follow what was done by \citet{Guainazzi:2006fk} using the regression method for left censoring data (\citealp{Schmitt:1985kx}, \citealp{Isobe:1986uq}). We carry out 10,000 Monte-Carlo simulations based on the results of the observations, applying the two following requirements: i) for the detections we substitute the values of the Fe K$\alpha$ EW with a random Gaussian distribution, in which the mean is the value obtained by the fit, and the standard deviation its error; ii) for the upper limits U we use a random uniform distribution in the interval [0,U]. For each Monte-Carlo run we fit the values with a log-linear relationship of the type
\begin{equation}\label{eq:xbe}
\log EW = \lambda + \mu \log L_{2-10}.
\end{equation}
We obtain an average slope $\overline{\mu}= -0.18\pm 0.06$, a value consistent with the results obtained by the recent studies of \citeauthor{Shu:2010zr} (\citeyear{Shu:2010zr}, $\mu=-0.13\pm0.04$) and \citeauthor{Shu:2012fk} (\citeyear{Shu:2012fk}, $\mu=-0.11\pm0.03$). In Fig.\,\ref{Fig:XBE} (left panel) we show the $EW-L_{2-10}$ scatter plot for our sample together with the best fit to the data.

An alternative approach to study the X-ray Baldwin effect is to use the ratio between the line and the continuum luminosity (i.e., $L_{\rm\,K\alpha}/L_{10-50}$) as a proxy of the EW. This allows to correct for the effects of absorption and to extend the study of the X-ray Baldwin effect to Seyfert\,2s. While the existence of a decrease of the relative flux of the Fe\,K$\alpha$ line with the luminosity has been found in several samples of Seyfert\,1s, studies of Seyfert\,2s have always taken into account the EW of the line, which is significantly enhanced for absorbed objects due to the damping of the continuum at $\sim 6-7$\,keV. We fit the {\it Suzaku} values of the different samples of Seyfert galaxies with the log-linear relation
 \begin{equation}\label{eq:xbe2}
\log \left(\frac{L_{\rm\,K\alpha}}{L_{10-50}}\right) =\phi + \omega \log L_{10-50}.
\end{equation}
To remove any bias introduced by absorption on the continuum luminosities of CT Sy2s and Sy2s, we use the same corrections applied in Sect.\,\ref{Sect:luminosities}. We find that the anti-correlation is statistically significant for both the Sy1s and Sy2s sample. For Sy1s we obtain a null hypothesis probability of $P_{\mathrm{n}}\simeq 0.2\%$ and a slope of $\omega_{\,1}=-0.11\pm0.01$. For Sy2s we find $P_{\mathrm{n}}\simeq 1\%$ and $\omega_{\,2}=-0.11\pm0.01$. To our knowledge this is the first statistically significant detection of the X-ray Baldwin effect in Seyfert\,2s. The fact that the slopes of Sy1s and Sy2s are the same is a strong indication that the mechanism responsible for the decrease of the relative flux of the iron K$\alpha$ line with the luminosity is the same. 
No significant trend is found considering only the sample of CT Sy2s. This is probably related to the absorption of part of the Fe\,K$\alpha$ flux, as demonstrated in Sect.\,\ref{Sect:absreflection}, which is bound to hide the anti-correlation. We tested whether the presence of radio-loud AGN might play a role in the correlation, and fitted the data excluding the four radio-loud type-I AGN, and the five radio-loud type-II. We found results in agreement with those obtained including these sources. In particular the slopes obtained are $\omega_{\,1}=-0.09\pm0.01$ ($P_{\mathrm{n}}\simeq 1.8\%$) and $\omega_{\,2}=-0.14\pm0.02$ ($P_{\mathrm{n}}\simeq 1.6\%$) for Sy1s and Sy2s, respectively.

\subsection{Comparison with time-averaged {\it Swift}/BAT data}\label{Sect:bat}

In order to reduce the possible effects of variability on the X-ray Baldwin effect, we combined the Fe\,K$\alpha$ luminosities of the {\it Suzaku} sample \citep{Fukazawa:2011fk} with 58-months averaged 14--195\,keV luminosity reported in the {\it Swift}/BAT catalog \citep{Baumgartner:2010uq}. We excluded the three sources not detected by {\it Swift}/BAT and the low-luminosity type-II AGN NGC\,4395, so that the final sample contained 30 Seyfert\,1s and 30 Compton-thin Seyfert\,2s. As proxy of the Fe\,K$\alpha$ EW we used the ratio between the luminosity of the Fe\,K$\alpha$ line and that of the 14--195\,keV continuum ($L_{\rm\,BAT}$). We did not find any statistically significant anti-correlation between the $L_{\rm\,K\alpha}/L_{\rm\,BAT}$ and $L_{\rm\,BAT}$. However, given the larger energy range of the BAT luminosities, the spread in photon indices might introduce a large scatter, and thus wash out the intrinsic trend. 
To remove any possible dependency of the luminosity on $\Gamma$ we calculated for each source, using the BAT photon indices, the monochromatic luminosity at 7.1\,keV. Using $L_{\rm\,7.1\,keV}$ instead of $L_{\rm\,BAT}$ we find an anti-correlation with slopes consistent, for both Sy1s ($\omega_{\,1}=-0.12\pm0.04$) and Sy2s ($\omega_{\,2}=-0.08\pm0.04$), with those obtained using {\it Suzaku} luminosities (Sect.\,\ref{sect:XBEtype1type2}), albeit with a larger scatter. This shows that variability does not play an important role in the anti-correlation.

\begin{figure}
\centering
\includegraphics[width=9cm]{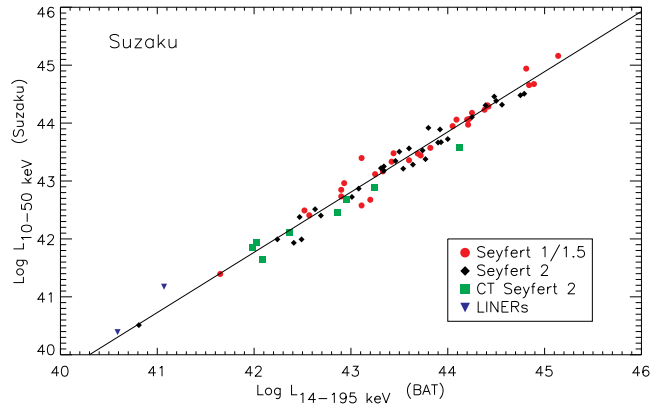}
 \caption{Scatter plot of the 10--50\,keV {\it Suzaku} luminosity versus the 14--195\,keV {\it Swift}/BAT luminosity. The continuous line represents the best linear fit to the data, with a slope of $K=1.04\pm0.2$.}
\label{Fig:lumsuzVSlumBat}
\end{figure}

In Fig.\,\ref{Fig:lumsuzVSlumBat} we illustrate the scatter plot of the the 10--50\,keV {\it Suzaku} luminosity versus the 14--195\,keV {\it Swift}/BAT luminosity. 
Performing a linear fit of the type  $\log L_{\rm\,10-50}\propto K \cdot \log L_{\rm\,14-195}$ on the data, we obtained $K=1.04\pm0.02$. The scatter in the correlation is mostly due to variability, and is similar in the two classes of objects: the standard deviation from the fit is of $\sigma=0.16$ and $\sigma=0.14$ for Sy1s and Sy2s, respectively.

\subsection{Explaining the anti-correlation}\label{sect:explainingXBE}
Several explanations to the X-ray Baldwin effect have been put forward since its discovery. \citet{Nandra:1997fk} and \citet{Nayakshin:2000uq} discussed how a luminosity-dependent ionisation state of the material where the line is produced might create the observed $EW-L_{\rm\,2-10}$ trend. Their models assumed that the Fe\,K$\alpha$ line is produced in the disk, but a similar scenario could be imagined for material located in the BLR or in the molecular torus. A luminosity-dependent ionisation state would imply a larger flux of ionised (H and He-like, e.g., \citealp{Matt:1996fk}) iron relative to neutral iron at high luminosities. However, \citet{Bianchi:2007vn} using a large sample of type-I AGN observed with {\it XMM-Newton} found only a weak statistically non-significant correlation between the ratio of the flux of ionised and neutral iron ($F(\mathrm{Fe\,XXV+Fe\,XXVI})/F(\mathrm{Fe\,I})$) and the X-ray luminosity. 

\cite{Jimenez-Bailon:2005tg} and \citet{Jiang:2006vn} argued that the correlation might be driven by the presence of radio-loud objects. However several recent works have showed that a significant correlation is also found when considering only radio-quiet AGN (e.g., \citealp{Bianchi:2007vn}, \citealp{Shu:2010zr}). Moreover, evidence of an X-ray Baldwin effect in radio-loud AGN has been found by \citet{Grandi:2006zr} using {\it BeppoSAX} observations.

\citet{Jiang:2006vn} proposed that the non-simultaneous reaction of the reprocessing material responsible for the Fe\,K$\alpha$ line to flux changes of the continuum might cause the trend. Given the distance of the torus from the X-ray source and its size, the line is in fact expected to react on much longer time-scales to flux changes of the continuum emission. This effect was investigated by \citet{Shu:2010zr}, who found that X-ray variability might cause at least part of the correlation, and that averaging different observations of the same source the slope of the X-ray Baldwin effect is flatter ($\mu=-0.13\pm0.04$) than the one obtained using all the observations ($\mu=-0.22\pm0.03$, see also Table\,1 in \citealp{Ricci:2013fk}). Another proof of the importance of variability was found by \citet{Shu:2012fk}, who detected a clear $EW-L_{\rm\,2-10}$ anti-correlation for individual sources. However, \citet{Shu:2012fk} carrying out simulations in which the X-ray continuum is variable and the Fe\,K$\alpha$ line is constant, found that variability cannot account for the whole X-ray Baldwin effect. Moreover, in Section\,\ref{Sect:bat} we have shown that, using X-ray luminosities averaged over several years, we still find evidence of an X-ray Baldwin effect both in Sy1s and Sy2s.

It has been shown that the Fe\,K$\alpha$ EW decreases with the Eddington ratio ($\lambda_{\rm\,Edd}$) with a slope similar to that obtained using the X-ray luminosity (e.g., \citealp{Bianchi:2007vn}). This leads to the possibility that the X-ray Baldwin effect might be related to the known relation between the photon index of the continuum power-law spectrum and the Eddington ratio (e.g., \citealp{Shemmer:2008fk}). The fact that for higher values of the Eddington ratio the photon index is steeper implies a lower number of photons at the energy of the Fe\,K$\alpha$ line, which leads to lower values of EW. In a recent paper \citep{Ricci:2013vn} we have carried out simulations to study the impact of this trend considering three different geometries of the reflecting material (toroidal, spherical toroidal and slab). We found that considering the $\Gamma-\lambda_{\rm\,Edd}$ relation obtained by \citeauthor{Shemmer:2008fk} (\citeyear{Shemmer:2008fk}; $\Gamma \propto 0.31 \log \lambda_{\rm\,Edd}$) one would obtain at most a slope of $\sim -0.08$, lower than the value found using the {\it Chandra}/HEG data of \citeauthor{Shu:2010zr} (\citeyear{Shu:2010zr}; $\log EW \propto (-0.13\pm 0.03)\log \lambda_{\rm\, Edd}$). A slope consistent with the observations would be produced using the results of \citet{Risaliti:2009nx} and \citet{Jin:2012fk}, who found a steeper increase of $\Gamma$ with $\lambda_{\rm\,Edd}$ ($\Gamma \propto 0.58 \log \lambda_{\rm\,Edd}$) than previous works. 
If the $\Gamma-\lambda_{\rm\,Edd}$ correlation is responsible for the X-ray Baldwin effect, then our results imply that the link between the X-ray corona and the accretion flow is similar in Sy1s and Sy2s.

A decrease of the fraction of obscured objects ($f_{\rm\,obs}$) with the luminosity has been observed in a very large number of surveys, from the radio to the hard X-rays (e.g., \citealp{Grimes:2004kx}, \citealp{Maiolino:2007bh}, \citealp{Simpson:2005uq}, \citealp{Beckmann:2009fk}). This phenomenon is often explained, in the framework of luminosity-dependent unification models, by the decrease of the covering factor of the torus with the luminosity, possibly caused by the effect of radiation pressure on the dust and gas of the torus. Because of the strong dependence of the Fe\,K$\alpha$ EW on the covering factor of the torus (e.g., \citealp{Ikeda:2009nx}), this mechanism has been often invoked as a possible explanation for the X-ray Baldwin effect (e.g., \citealp{Page:2004kx}, \citealp{Zhou:2005ys}). In a recent work \citep{Ricci:2013fk}, we have used physical torus models and recent X-ray and IR measurements of the decrease of $f_{\rm\,obs}$ with the luminosity to quantify the effect of this phenomenon on the Fe\,K$\alpha$ EW. We found that in the Seyfert regime ($42 \leq \log L_{\,2-10} \leq 44.2$) luminosity-dependent unification can reproduce the slope of the X-ray Baldwin effect for a large range of values of equatorial column densities of the torus ($\log N_{\rm\,H}^{\rm\,T} \geq 23.1$). In the quasar regime ($ \log L_{\,2-10} > 44.2$) the situation is less clear, due to the smaller sample and to the large errors associated to the values of the Fe\,K$\alpha$ EW, but our results seem to indicate that a slower decrease of $f_{\rm\,obs}$ with $ L_{\,2-10}$ than that observed at lower luminosities is necessary to explain the observations. In the Seyfert regime the intercept of the X-ray Baldwin effect found by {\it Chandra}/HEG observations \citep{Shu:2010zr} can be explained only by a torus with an equatorial column density of $\log N_{\rm\,H}^{\rm\,T} \simeq 23.2$. This value is lower than the $N_{\rm\,H}$ observed in many Sy2s, and might be related to the fact that the spherical-toroidal geometry we used is expected to produce values of EW larger than a toroidal geometry for the same set of parameters. If the decrease of the covering factor of the torus with the luminosity is the cause of the X-ray Baldwin effect, our findings show that on average the tori of Sy2s behave similarly to those of Sy1s.

Amongst the different explanations proposed for the X-ray Baldwin effect, so far luminosity-dependent unification seems to be the most promising. The effect of the $\Gamma-\lambda_{\rm\,Edd}$ trend would also be able to account for the observed $EW-L_{\,2-10}$ relation, but only for a steep increase of $\Gamma$ with $\lambda_{\rm\,Edd}$, which has yet to be confirmed. In order to discern between these two explanations it is necessary to understand whether the main driver of the X-ray Baldwin effect is the luminosity or the Eddington ratio. In \citet{Ricci:2013vn} we show that the $EW-L_{\,2-10}$ anti-correlation is statistically more significant than the $EW-\lambda_{\rm\,Edd}$ one. However uncertainties on the estimates of the black hole masses and on the bolometric corrections do not allow to fully discard the possibility that $\lambda_{\rm\,Edd}$ is the main driver of the X-ray Baldwin effect.

\section{Summary and conclusions}\label{Sect:summary}
In this work, we studied the relation between the narrow Fe\,K$\alpha$ line and the X-ray continuum in different types of AGN, using the results of several X-ray works carried out using {\it Chandra}, {\it XMM-Newton}, and {\it Suzaku} . Our main results are the following.
\begin{itemize}

\smallskip
\item The luminosity of the Fe\,K$\alpha$ line normalised to that of the continuum is on average lower in Sy2s than in Sy1s (right panel of Fig.\,\ref{Fig:LcLkalpha}). Studying the average values of $L_{\rm\,K\alpha}/L_{10-50}$ obtained simulating dummy Sy1s and Sy2s populations, we showed that this difference is consistent with being due to different average inclination angles with respect to the molecular torus (left panel of Fig.\,\ref{Fig:expected_ratio}), confirming the existence of an axisymmetric structure (torus-like) responsible for the reprocessing of the X-ray radiation. Alternatively, this difference might be due to differences in the intensities of Compton humps, in the photon index distributions or in the average iron abundances.

\smallskip
\item We showed that the ratio between the flux of the broad and narrow Fe K$\alpha$ line in the 6.35--6.45\,keV range depends on the torus geometry considered, and is on average $<25\%$ and $<15\%$ for type\,I and type\,II AGN, respectively (see Figs.\,\ref{Fig:ratio_broadnarrow} and \ref{Fig:ratio_broadnarrow_torusmod}).

\smallskip
\item We showed that the luminosity of the Fe\,K$\alpha$ line is attenuated in CT Sy2s (Fig.\,\ref{Fig:ratiovsNH}) due to absorption of part of the reflected component. This result is also confirmed comparing the $L_{\rm\,K\alpha}$ of the AGN to their MIR luminosity (Fig.\,\ref{Fig:lumMIR} and Table\,\ref{tab:fit_MIR}), which shows that the luminosity of the Fe\,K$\alpha$ line in CT Sy2s is systematically lower than that of Sy1s and Sy2s for the same MIR luminosity. This implies that i) the Fe\,K$\alpha$ luminosity cannot be used to estimate the intrinsic bolometric luminosity of AGN, ii) care should be taken when studying the luminosity function of CT Sy2s, iii) the X-ray spectral analysis of obscured type-II AGN should also take into account absorption of the reflection component. We estimated that on average, in Compton-thin and CT Sy2s, the reflected radiation is seen through a column density which is about $1/4$ of the one absorbing the primary X-ray emission.

\smallskip
\item We found the first significant evidence of X-ray Baldwin effect in Seyfert\,2s. Our study has shown that for Sy2s $L_{\rm\,K\alpha}/L_{10-50}$ decreases with $L_{10-50}$ with the same slope observed in Sy1s (right panel of Fig.\,\ref{Fig:XBE}). This implies that the mechanism responsible for this effect is the same in the two classes of AGN. If the mechanism responsible for the X-ray Baldwin effect is the $\Gamma-\lambda_{\rm\,Edd}$ correlation, then the X-ray corona and the accretion flow are connected in a similar way in Sy1s and Sy2s. On the other hand, if the X-ray Baldwin effect is due to luminosity-dependent unification of AGN, then our results show that the decrease of the covering factor of the torus in Sy2s is similar to that of Sy1s. 

\end{itemize}

\appendix

\section{Suzaku data}\label{list_sources}
In Table\,\ref{tab:data1} the values of the 10--50\,keV and Fe K$\alpha$ luminosities for the LINERs and type-I AGN of the {\it Suzaku} sample of \cite{Fukazawa:2011fk} are listed. Table\,\ref{tab:data2} reports the same properties of Table\,\ref{tab:data1} for the type-II and CT AGN, plus the absorption-corrected luminosities. The approach adopted to correct the luminosities is reported in Sect.\,\ref{Sect:Sy1vsSy2}.

\begin{table}
\caption[]{10--50\,keV (1) and Fe K$\alpha$ (2) luminosities for the LINERs and type-I AGN of the {\it Suzaku} sample of \cite{Fukazawa:2011fk}.}
\label{tab:data1}
\begin{center}
\begin{tabular}{lccc}
\hline
\hline
\noalign{\smallskip}
& (1) & (2)     \\ 
\noalign{\smallskip}
Source 						  & $\log L_{\,10-50}$  & $\log L_{\rm\,K\alpha}$       \\ 
\noalign{\smallskip}
\hline
\noalign{\smallskip}
\noalign{\smallskip}

\multicolumn{3}{c}{\textbf{LINERs}} \\
\noalign{\smallskip}
\noalign{\smallskip}

M81 &   40.39 &37.92	 	\\
M106 &     41.17  &38.67       		\\
\noalign{\smallskip}
\noalign{\smallskip}

\multicolumn{3}{c}{\textbf{Type-I AGN}} \\
\noalign{\smallskip}
\noalign{\smallskip}

1H\,0419$-$577 		   &  44.94 &42.10	\\
3C\,120 				   &  44.23 &41.80		\\
3C\,382 				   &  44.66 &42.12		\\
3C\,390.3 				    &  44.67 &42.12		\\
4C\,+74.26 			   &  45.16 &42.58		\\
ARK\,120 				   &  44.06 &41.76		\\
Fairall\,9 				   &  44.29 &42.14	\\
IC\,4329A 				   &  44.07 &41.55		\\
IGR\,J16185$-$5928	 	   &  43.57 &41.26			\\
IGR\,J21247+5058 		   &  44.18 &41.23		\\
IRAS\,18325$-$5926 	   &  43.40 &40.95	\\	 
MCG+8$-$11$-$11 		   &  44.06 &41.75	\\
MCG$-$6$-$30$-$15	   &  42.96 &40.42	\\
Mrk\,79 				   &  43.44 &41.39		\\
Mrk\,110 				   &  43.97 &41.48	\\
Mrk\,335 				   &  43.48 &41.10		\\
Mrk\,359 				   &  42.73 &40.62		\\
Mrk\,509 				   &  44.31 &41.87		\\
Mrk\,766 				   &  42.85 &40.43	\\
Mrk\,841 				   &  43.95 &41.51	\\
NGC\,3227 			   &  42.41 &40.12	\\
NGC\,3516 			   &  43.17 &40.99	\\
NGC\,3783 			   &  43.36 &41.20	\\
NGC\,4051 			   &  41.39 &39.14		\\
NGC\,4151 			   &  42.58 &40.53		\\
NGC\,4593 			   &  42.68 &40.68		\\
NGC\,5548 			   &  43.47 &41.14		\\
NGC\,7213 			     &  42.49 &40.22		\\
Swift\,J0501.9$-$3239 	     &  43.12 &40.73		\\		 
Swift\,J2009.0$-$6103 	     &  43.33 &40.71		\\
\noalign{\smallskip}
\hline
\noalign{\smallskip}
\end{tabular}        
\end{center}
\end{table}

\begin{table}
\caption[]{Observed (1) and absorption corrected (2) 10--50\,keV luminosities, and Fe K$\alpha$ luminosities (3) for the Type-II and CT AGN of the sample of \cite{Fukazawa:2011fk}.}
\label{tab:data2}
\begin{center}
\begin{tabular}{lccc}
\hline
\hline
\noalign{\smallskip}
& (1) & (2) & (3)      \\ 
\noalign{\smallskip}
Source 						  & $\log L^{\rm\,obs}_{\,10-50}$& $\log L_{\,10-50}$  & $\log L_{\rm\,K\alpha}$       \\ 
\noalign{\smallskip}
\hline
\noalign{\smallskip}

\noalign{\smallskip}
\multicolumn{4}{c}{\textbf{Type-II AGN}} \\
\noalign{\smallskip}
\noalign{\smallskip}

3C\,33 							&   44.31		&      44.51    	&      41.86  					\\		
3C\,105 							&   44.51		&      44.66    	&      42.18  						\\		 			
3C\,445 							&   44.38		&      44.50    	&      41.99  									\\ 			
3C\,452  							&   44.48		&      44.68    	&      42.05  									\\			
Centaurus\,A 						&   42.72		&      42.78    	&      40.34  									\\	 	
ESO\,263-G013 					&   43.67		&      43.76    	&      41.14  								\\ 	
MCG+04$-$48$-$002 				&   43.21		&      43.46    	&      40.59  								\\ 	
MCG$-$5$-$23$-$16 				&   43.51		&      43.63    	&      41.17  								\\ 	
Mrk\,3  							&   43.53		&      43.80    	&      41.30  								\\			
Mrk\,417  							&   43.89		&      44.06    	&      41.28  								\\			
Mrk\,1210 						&   43.18		&      43.31    	&      40.80  									\\ 		
NGC\,1052 						&   41.99		&      42.09    	&      39.70  									\\ 		
NGC\,1365 						&   42.51		&      42.55    	&      39.95  									\\ 		
NGC\,2110 	 					&   43.56		&      43.62    	&      41.02  								\\	
NGC\,3281 						&   43.25		&      43.48    	&      40.87  									\\ 		
NGC\,4388 						&   43.28		&      43.42    	&      41.08  									\\ 		
NGC\,4395 						&   40.51		&      40.52    	&      37.95  									\\ 		
NGC\,4507 						&   43.38		&      43.60    	&      41.24  									\\ 		
NGC\,4992 						&   43.66		&      43.84    	&      41.30  									\\ 		
NGC\,5506	 					&   43.22		&      43.24    	&      40.82  								\\	
NGC\,6300 						&   42.38		&      42.46    	&      39.91  								\\ 		
NGC\,7172 						&   43.35		&      43.40    	&      40.79  								\\ 		
NGC\,7314 						&   41.93		&      41.94    	&      39.83  									\\ 		
NGC\,7582						&   42.40		&      42.60    	&      40.08  									\\	
Swift\,J0138.6$-$4001 				&   43.72		&      43.93    	&      41.35  									\\ 	
Swift\,J0318.7+6828 				&   44.32		&      44.38    	&      41.90  								\\ 	
Swift\,J0505.7$-$2348				&   44.10		&      44.17    	&      41.60  									\\ 	
Swift\,J0601.9$-$8636				&   41.99		&      42.25    	&      40.25  									\\ 	
Swift\,J0959.5$-$2258				&   42.87		&      43.15    	&      40.55  								\\ 	
Swift\,J1200.8+0650 				&   43.92		&      43.99    	&      41.33  								\\ 	
Swift\,J1628.1+5145 				&   44.46		&      44.53    	&      41.76  									\\ 	
Mrk\,1239 						&   42.84		&      43.01    	&      40.46  									\\ 		
NGC\,2273 						&   41.92		&      42.22    	&      40.26  								\\ 		
NGC\,1142 						&   43.92  		&      44.14    	&      41.54    									\\			

\noalign{\smallskip}

\noalign{\smallskip}
\multicolumn{4}{c}{\textbf{CT AGN}} \\
\noalign{\smallskip}
\noalign{\smallskip}

Circinus\,Galaxy				&   	41.65		&    42.01  		&        39.88			  						\\
ESO\,323-G032 				&   	42.46		&    42.86  		&        40.56									\\
ESO\,506-G027 				&   	43.58		&    43.90  		&        41.48								\\
IRAS\,19254-7245 				&   	44.05		&    44.99  		&        40.98								\\
Mrk\,1073 					&   	42.90		&    43.59  		&        40.63									\\
Mrk\,273 						&   	43.46		&    43.87  		&        41.04								\\
NGC\,1068 					&   	41.94		&    42.28  		&        40.28									\\
NGC\,1386 					&   	41.28		&    41.90  		&        39.06										\\
NGC\,3079 					&   	41.86		&    42.50  		&        39.15									\\
NGC\,3393 					&   	42.68		&    43.02  		&        40.22									\\
NGC\,4945 					&   	42.11		&    42.84  		&        39.47									\\
NGC\,5135 					&   	42.74		&    43.39  		&        40.57								\\
NGC\,5728 					&   	42.90		&    43.27  		&        40.58									\\

\noalign{\smallskip}
\hline
\noalign{\smallskip}
\end{tabular}        
\end{center}
\end{table}

\section*{Acknowledgments}
We thank the referee for his/her comments, that significantly helped to improve the quality of the manuscript.
We thank Chin Shin Chang for her comments on the manuscript. CR is a Fellow of the Japan Society for the Promotion of Science (JSPS). This work was partly supported by the Grant-in-Aid for Scientific Research 23540265 (YU) from the Ministry of Education, Culture, Sports, Science and Technology of Japan (MEXT). PG acknowledges support from STFC grant reference ST/J00369711. This research has made use of the NASA/IPAC Extragalactic Database (NED) which is operated by the Jet Propulsion Laboratory, of data obtained from the High Energy Astrophysics Science Archive Research Center (HEASARC), provided by NASA's Goddard Space Flight Center, and of the SIMBAD Astronomical Database which is operated by the Centre de Donn\'ees astronomiques de Strasbourg.

\clearpage
\bibliographystyle{mnras}
\bibliography{iron_paper_2}

\end{document}